 \documentclass[sigconf]{acmart}
 
\AtBeginDocument{%
  }

\copyrightyear{2026}
\acmYear{2026}
\setcopyright{cc}
\setcctype{by}
\acmConference[CHI '26]{Proceedings of the 2026 CHI Conference on Human Factors in Computing Systems}{April 13--17, 2026}{Barcelona, Spain}
\acmBooktitle{Proceedings of the 2026 CHI Conference on Human Factors in Computing Systems (CHI '26), April 13--17, 2026, Barcelona, Spain}
\acmPrice{}
\acmDOI{10.1145/3772318.3791977}
\acmISBN{979-8-4007-2278-3/2026/04}

\acmSubmissionID{1614}

\usepackage{multirow}
\usepackage{enumitem}
\usepackage{siunitx}
\usepackage{siunitx}
\usepackage{graphicx}  
\usepackage{soul}
\usepackage{float}
\usepackage{subcaption}

\begin{document}

\title [What Are You Really Asking For?]
{What Are You Really Asking For? A Comparative 5W1H Analysis of Learner Questioning in CPR Training with IVAs in Screen-based and Augmented Reality Environments}


\author{Hyerim Park}
\orcid{0000-0001-6764-4490}
\affiliation{%
\institution{Post Metaverse Research Center} 
  \institution{KAIST} 
  \city{Daejeon}
  \country{Republic of Korea}
}
  \email{ilihot@kaist.ac.kr}

\author{Jinseok Hong}
\orcid{0009-0008-4647-5016}
\affiliation{%
  \institution{UVR Lab}
    \institution{KAIST} 
\city{Daejeon}
  \country{Republic of Korea}}
\email{jindogliani@kaist.ac.kr}

\author{Heejeong Ko}
\orcid{0000-0001-6898-4104}
\affiliation{%
  \institution{UVR Lab}
    \institution{KAIST} 
    \city{Daejeon}
  \country{Republic of Korea}}
\email{hj.ko@kaist.ac.kr}

\author{Woontack Woo}
\orcid{0000-0002-5501-4421}
\authornote{Corresponding author.}
\affiliation{%
  \institution{UVR Lab and KI-ITC ARRC}
    \institution{KAIST} 
  \city{Daejeon}
  \country{Republic of Korea}}
\email{wwoo@kaist.ac.kr}






\renewcommand{\shortauthors}{Park et al.}


\begin{abstract}
Question-asking is one of the key indicators of cognitive engagement. However, understanding how the distinct psychological affordances of presentation media shape learners' spoken inquiries with embodied Intelligent Virtual Agents (IVAs) remains limited. To systematically examine this process, we propose a 5W1H-based framework for analyzing learner questions.
Using this framework, we conducted a user study comparing an Augmented Reality-based IVA (AR-IVA) deployed in the physical environment with a screen-based IVA (Video-IVA) during cardiopulmonary resuscitation (CPR) instruction. Results showed that the AR-IVA elicited higher spatial and social presence and promoted more frequent and longer questions focused on clarification and understanding. In contrast, the Video-IVA encouraged questions regarding procedural refinement. Presence acted as a selective filter, shaping the timing and topic of questions rather than as a universal mediator. These effects were significantly moderated by learners’ motivational and strategic characteristics toward learning. Based on these findings, we propose design implications for IVA-supported learning systems.

\end{abstract}






\begin{teaserfigure}
  \includegraphics[width=\textwidth]{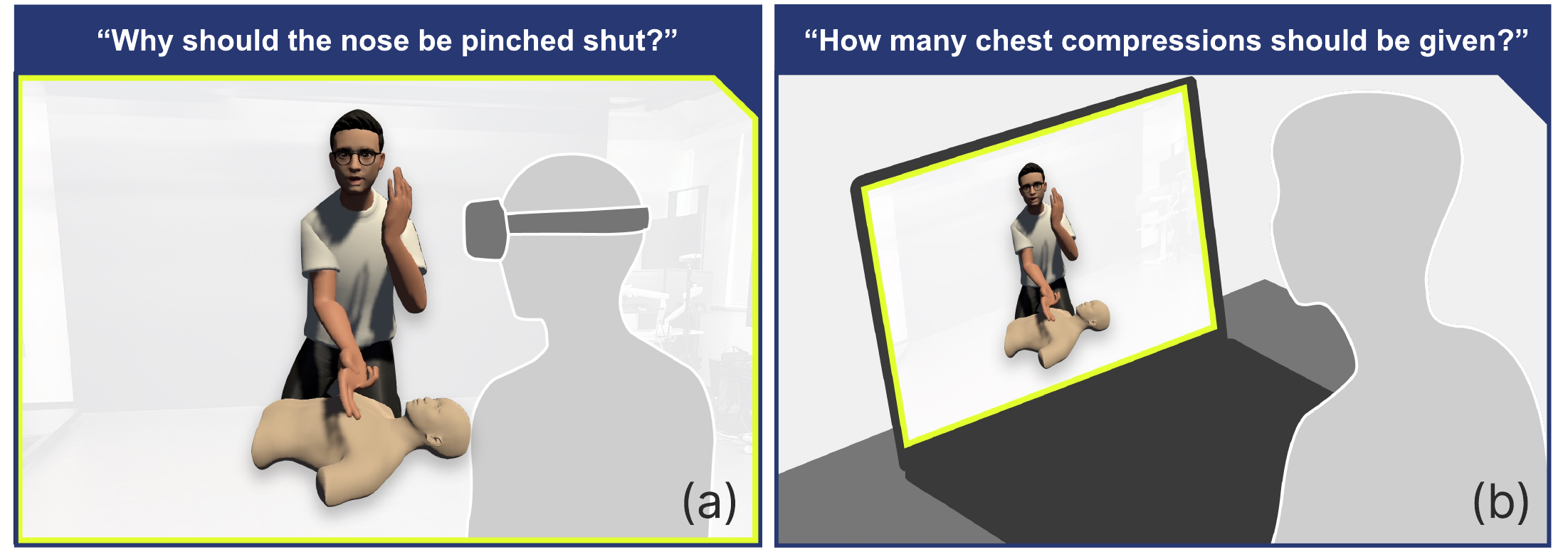}
  \caption{Comparison of learner questioning behavior with Intelligent Virtual Agents (IVAs) presented in two modalities: (a) augmented reality (AR) environment, where learners experienced higher spatial and social presence and asked conceptual-deepening questions; (b) screen-based (Video) environment, where learners engaged with the same IVA on a flat display and generated procedural-refinement questions.}
  \Description{Figure (a) shows a conceptual illustration of the AR-IVA environment, where a learner (represented by the silhouette wearing an HMD) interacts with a 3D virtual instructor performing CPR on a virtual mannequin. In this condition, the phrase "Why should the nose be pinched shut?" is displayed as a sample inquiry, reflecting the learners' tendency to pursue conceptual deepening by probing underlying principles. Figure (b) represents the Video-based IVA modality, where the learner interacts with the virtual instructor through a 2D screen interface. In this setting, the text "How many chest compressions should be given?" is presented as a sample inquiry, illustrating the learners' focus on procedural refinement to verify the accuracy and steps of the protocol.}
  \label{fig:teaser}
\end{teaserfigure}


\maketitle

\section{Introduction}

Question-asking serves as a key indicator of learners' cognitive engagement, directly reflecting how they interpret, explore, and construct knowledge from instructional content~\cite{king1994guiding, graesser1994question,chin2008students}. With the widespread adoption of Large Language Model (LLM)–based conversational services, such as ChatGPT and Gemini, research has extensively investigated how learners interact with AI partners to facilitate learning~\cite{kasneci2023chatgpt,memarian2023chatgpt}. 
Prior studies have examined learners’ prompt patterns and dialogue logs to uncover information-seeking strategies~\cite{kim2024using,viberg2025chatting}, metacognitive regulation~\cite{fan2025enhancing,hao2026understanding}, and conceptual understanding processes~\cite{anthony2004student,chi2014icap}. However, these analytical insights remain largely restricted to text-based chat interfaces, limiting our understanding of how questioning unfolds in more naturalistic, embodied contexts.

LLMs have recently been extended into avatar-based Intelligent Virtual Agents (IVAs), enabling learners to interact with agents through natural, spoken dialogue~\cite{embodiedagentsinXR, zhu2025pedagogical, tan2025aichatbot}. These embodied IVAs provide high visual fidelity and heightened social presence, offering interaction experiences akin to human tutoring. Despite these technological advancements, research has largely concentrated on system-level performance~\cite{buldu2025cuify,maslych2025mitigating} and user-experience outcomes, such as presence, usability, and learning performance~\cite{zhang2024virtual,hassan2022towards,nam2025effects}. Little attention has been paid to the process of how learners construct and adapt their spoken inquiries during interaction.
In particular, systematic comparative research is lacking on how the psychological affordances of educational media~\cite{makransky2021cognitive} and learner characteristics jointly shape learners’ questioning behavior.

Therefore, we propose a 5W1H-based framework to systematically analyze learners’ questioning behavior with avatar-based IVAs. The framework captures \textit{who} asks, \textit{what} is asked, \textit{when}, \textit{where} and \textit{why} questions arise, and \textit{how} they are expressed. We implemented a real-time question-and-answer (Q\&A) IVA system delivering Cardiopulmonary Resuscitation (CPR) instruction. In the user study, the identical IVA was presented in two presentation modes: an Augmented Reality-based (AR-IVA)  and a screen-based (Video-IVA) mode.
Forty participants were assigned to one of these conditions and encouraged to ask questions freely during and after instruction. All spoken inquiries were automatically transcribed and logged. Using this data, we examined the direct effects of presentation modality on psychological affordances (specifically spatial and social presence), questioning behavior, and learning outcomes. Subsequently, we analyzed the mediating role of spatial and social presence and the moderating role of learners' motivation and strategies in the influence of presentation modality on questioning behavior.

Our results show that learning with AR-IVA increased spatial and social presence, improved learning performance, and significantly altered questioning behavior compared to Video-IVA. Specifically, the AR-IVA promoted more frequent and longer questions oriented toward conceptual deepening, whereas the Video-IVA encouraged procedural refinement. However, these behavioral changes were not directly mediated by the presence. Instead, heightened presence acted as a selective filter, minimizing disruptive questioning during the demonstration stage and focusing attention on the conceptual core. 
We also found that variations in questioning timing and content—such as topic, type, and form—were driven primarily by learners’ motivation and strategies rather than by the medium itself. These results suggest that learner questioning behavior with avatar-based IVAs is not determined solely by the technological medium but emerges through complex interactions with individual learner characteristics. Our study contributes to the understanding of avatar-based IVA interaction in three key ways.

\begin{itemize} 

    \item First, we propose a 5W1H-based framework to systematically analyze learners' questioning behavior during interactions with avatar-based IVAs. 
    \item Second, we empirically demonstrate how questioning behavior is shaped by the interplay between the psychological affordances of the presentation modality and learner traits. 
    \item Finally, we derive design implications for avatar-based IVA systems to enhance learning by integrating presentation modalities with learners’ traits.
    
\end{itemize}

\section{Related Work} 

\subsection{LLM-Driven Virtual Agents: From Generation to Experience}
Research integrating LLMs into AR and VR (virtual reality) environments has primarily centered on enhancing systems’ generative capabilities, particularly the quality of agent outputs~\cite{tang2025llm}. 
In educational contexts, technical efforts have focused on developing virtual agents that generate context-aware dialogue using LLMs. 
For instance, virtual agents combining GPT-4 with augmented context retrieval have been developed to serve as laboratory assistants in VR environments~\cite{ayre2023implementation}. Additionally, AR language tutors integrating computer vision with LLMs have been proposed to provide context-aware dialogue and real-time feedback~\cite{lee2023visionary}. Furthermore, integrating LLMs into virtual robots has enabled natural social interactions within educational game scenarios~\cite{bottega2023jubileo}. Beyond education, research has also integrated spatial computing or AR interfaces with LLMs to develop agents that provide targeted guidance across diverse contexts~\cite{izquierdo2023large, spiegel2024feasibility}. Collectively, these studies have focused on advancing and evaluating the expressive and communicative capabilities of LLM-driven virtual agents.

In contrast, user-centered research has emphasized experiential outcomes, primarily measured through questionnaires~\cite{embodiedagentsinXR}. These include system usability~\cite{reinhardt2020embedding} and technology acceptance~\cite{safadel2023user}, as well as perceptions of virtual agents (e.g., attractiveness, trust, and knowledgeability)~\cite{yang2024effects, zhu2024virtual, yang2025exploring}. User experience metrics such as spatial and social presence~\cite{pan2025ellma, elfleet2024investigating}, along with psychological assessments related to anxiety or stress~\cite{fang2025social}, have also been examined. Additional quantitative measures, such as interaction duration~\cite{gan2023design}, number of speech turns~\cite{llanes2024developing}, or heart rate~\cite{feli2025llm}, have been employed alongside qualitative interviews and thematic analyses.

However, the inquiry process, specifically how learners construct and adapt questions while interacting with an agent, has been largely overlooked. Furthermore, there is a lack of systematic comparative studies investigating how questioning behavior varies depending on the affordances of the medium through which the IVA is delivered or on learner characteristics. Accordingly, this study examines how avatar-based IVAs shape learners’ questioning behavior across media affordances and learner characteristics.



\subsection{Determinants of Questioning: Media Affordances and Learner Traits}

To understand how avatar-based IVAs shape questioning behavior, it is essential to examine the factors that influence information seeking: the psychological affordances of the presentation modality and learner characteristics.

\subsubsection{Media Affordances}
According to the Cognitive Affective Model of Immersive Learning (CAMIL), the psychological affordances of immersive media, such as spatial presence, social presence, and agency, are critical drivers of the learning experience~\cite{makransky2021cognitive}. Spatial presence strengthens situational grounding, enabling learners to interpret instructional cues as if they occupy the real environment. Social presence, enhanced by embodied virtual agents, fosters co-presence and dialogue patterns akin to human tutoring. Agency reflects learners’ perceived ability to control actions in the environment, shaping how they engage with instructional content. While these affordances are known to influence motivation, cognitive load, and learning outcomes both directly and indirectly~\cite{kuhne2023direct,arjunamahanthi2025interactive,johnson2021platform}, their role in shaping the information-seeking process remains obscure.

\subsubsection{Learner Motivation and Self-Regulated Strategies}
Questioning behavior reflects learners' internal attributes, such as domain knowledge, motivation, and Self-Regulated Learning (SRL) strategies, rather than external conditions alone~\cite{king1994guiding, graesser1994question,newman2023adaptive,faza2025self}. From a social-cognitive perspective, motivation and learning strategies are dynamic, context-sensitive constructs rather than fixed traits. Specifically, motivation comprises three key components: value beliefs (e.g., intrinsic or extrinsic goal orientations),
expectancy beliefs (e.g., self-efficacy and control of learning), and affective factors (e.g., test anxiety).
Similarly, learning strategies encompass tactics under the learner's control, categorized into three domains: cognitive strategies (e.g., elaboration and organization), metacognitive strategies (e.g., planning and monitoring), and resource-management strategies (e.g., time management and help-seeking)~\cite{duncan2005making}.

Recent research shows that external scaffolds, such as AI-mediated support and adaptive feedback systems, interact with these internal characteristics, actively eliciting specific regulatory behaviors~\cite{maimaiti2025gamified,faza2025self}. Learners also adjust their help-seeking behaviors by aligning their internal goals with the affordances provided by the system~\cite{chen2025unpacking,fan2025beware,karumbaiah2022context}. Thus, understanding how IVA presentation modality interacts with learner motivation and strategies is essential for designing systems that support strategic inquiry.

\subsection{Adapting the 5W1H Framework for Embodied Interaction}
The analysis of learner questioning has relied on form and intent classification~\cite{madabushi2016questionclassification, banerjee2012questionclassification}, topic modeling using queries and click graphs~\cite{JANSEN20081251, craswell2007clickgraph}, and query-volume analysis as a proxy for engagement~\cite{CHEN201821, ghazarian2020}.
With the advent of LLMs, these techniques have expanded to include LLM-assisted intent classification, semantic topic induction, and query rewriting or expansion~\cite{kim2024using, ALFARABY2024100298, wang2023query2doc}, enabling deeper inferences about learners’ information-seeking processes.
However, they often fail to capture the multifaceted context of immersive environments, where spatial cues, temporal dynamics, and learner traits are inextricably linked.

Given the complex interplay between the psychological affordances of presentation modality and learner characteristics, described above, a unified analytical framework is required to systematically capture this multidimensional structure. The 5W1H framework (Who, When, Where, What, Why, How) has been widely established as a robust model for analyzing user behavior and situational context in ubiquitous computing and context-aware systems~\cite{liang2025physical,yu2010context,jang2005unified}. In the domain of Human-AI Interaction, this framework has been utilized to classify user needs into functional and emotional dimensions for LLM application design~\cite{chen2024designfusion} and to categorize engagement modes~\cite{gao2024taxonomy}.
It has also been applied to structurally analyze qualitative log data from unstructured LLM interactions~\cite{zhu2025data}. These studies illustrate how the framework not only structures interaction patterns but also informs the design of LLM-based systems by clarifying how and why users engage with AI. 

Building on these perspectives, this study adapts the 5W1H framework to bridge the analytical gap between text-based and embodied voice-based interactions. This framework enables a comprehensive analysis that goes beyond simple frequency counts to examine \textit{Who} asks, \textit{When} and \textit{Where} questions arise, \textit{What} and \textit{Why} they ask, and \textit{How} questions are expressed. By applying this integrated framework, we aim to clarify how avatar-based IVAs shape the procedural and strategic patterns of learner questioning.

\section{Method}
This section details our methodological approach to investigating learner questioning behavior with avatar-based IVAs. First, we establish a 5W1H-based framework to systematically characterize questioning behavior. Based on this framework, we formulate our research hypotheses. Subsequently, we describe the development of the LLM-driven IVA system, the experimental setup, and the user study procedure. We then specify how questioning behavior is operationalized within the 5W1H-based framework and describe the dependent variables used to capture learners’ experiences. Finally, we outline the data analysis strategy employed to evaluate our hypotheses.

\subsection{Questioning Behavior Analysis Framework}
\label{sec:Framework}

\begin{table*}[t]
\centering
\caption{5W1H-based framework for defining and measuring learner questioning behavior.}
\label{tab:questioning_framework}
\resizebox{.96\linewidth}{!}{%
\begin{tabular}{@{}lll@{}}
\toprule
\textbf{Dimension} & \textbf{Definition} & \textbf{Measures} \\ 
\midrule
Who   & Learner traits and characteristics \emph{(Who asked the question?)}  & Motivation, strategies \\
When  & Temporal context of questioning \emph{(When was the question asked?)}  & Absolute and contextual time position \\
Where & Spatial focus at the moment of questioning \emph{(Where was the learner looking?)} & Eye-tracking \\
What & Semantic content of the question \emph{(What was the learner asking about?)} & Topic, type, form \\
Why  & Underlying learner intention of the question \emph{(Why did the learner ask?)} & Intention labels\\
How   & Intensity and depth of the question \emph{(How was the question expressed?)} & Volume, complexity \\
\bottomrule
\end{tabular}%
}
\end{table*}

During interactions with avatar-based IVAs, merely counting or categorizing questions at a surface level is insufficient to capture the complexity and richness of questioning behavior.
Therefore, a more systematic and multidimensional analysis is required to understand how this questioning behavior unfolds within specific educational contexts and modalities.
To address this need,  as shown in Table~\ref{tab:questioning_framework}, we propose a 5W1H-based framework that disaggregates questioning into six complementary dimensions: Who (learner traits), When (temporal position), Where (attentional
target), What (content focus), Why (intention), and How (intensity and depth). In the following, we detail how each dimension was defined and measured.

\textbf{Who. } 
This dimension focuses on intrinsic learner characteristics that shape questioning behavior. It is operationalized through two factors: learning motivation and learning strategies. Learning motivation serves as an internal driver of inquiry, influencing a learner’s likelihood to ask questions. Learning strategies function as cognitive tools that learners employ to organize and process information, shaping how questions are formulated. These characteristics are critical for understanding individual variability in question-asking and for informing personalized interaction design.

\textbf{When. }
This dimension captures the temporal aspect of questioning and is measured using two indicators: absolute time position and contextual time position.
The absolute time position records the precise timestamp of each question, while the contextual time position identifies the instructional stage or event occurring at that moment.
These indicators allow us to determine when questions arise during instruction and how they align with the progression of learning stages.
These temporal patterns reveal how presentation modalities and learner characteristics influence question timing, thereby enabling the identification of instructional segments that may require additional scaffolding in IVA design. 

\textbf{Where. }
This dimension captures the learner’s visual attention at the moment of questioning.
It is measured using eye-tracking to identify the precise object or location of the learner’s gaze when the question is asked.
This approach establishes a direct link between questioning behavior and visual focus, showing how learners connect where they see with what they ask.
Such insights support the design of IVAs that adapt cues and explanations to shifts in learners’ visual attention, while proactively guiding focus toward key instructional elements.

\textbf{What. }
This dimension examines the content and structure of questions through three aspects: topic, type, and form.
Topic identifies the specific subject matter, revealing the content on which learners concentrate.
Type represents the nature of the knowledge the learner seeks. Procedural inquiries focus on how to perform specific actions, while exploratory inquiries aim to understand underlying concepts or the rationale behind a process.
Form describes the linguistic structure (e.g., what, why, how), indicating how learners frame their inquiries.
These aspects clarify the specific knowledge learners aim to acquire and reveal how different modalities influence the nature of their inquiries.

\textbf{Why. }
This dimension captures the motivational and informational intention underlying each learner’s question. This intention can be identified using various approaches, ranging from automated methods—such as supervised classification, unsupervised pattern discovery, or LLM-based inference—to manual annotation. By systematically analyzing learner intentions, we can support the design of IVAs capable of interpreting user goals. This enables the system to tailor responses that better align with learners’ needs, accounting for both media conditions and individual characteristics.

\textbf{How. }
This dimension captures the expressive intensity and linguistic depth of a learner’s questioning behavior. Intensity is reflected in the overall volume of questions generated, while depth is assessed via their linguistic complexity. These indicators reflect the learner’s engagement and cognitive progress, enabling the IVA to infer the learner’s state and tailor its responses accordingly.

Note on the ``Where'' dimension.
Although the framework includes a spatial dimension (``Where''), this component was not analyzed in the present study. Because the experimental conditions presented the same instructional elements and visual focal points, learners’ attentional targets did not vary meaningfully across conditions. As such, the ``Where'' dimension was not included in comparative analyses but remains an integral part of the broader framework for future applications where spatial variation is present.


\subsection{Hypotheses}
To analyze the differential effects of IVA presentation modality (AR-IVA vs. Video-IVA) on learner experience and questioning behavior, we formulated the following research hypotheses based on the 5W1H-based framework. Specifically, we hypothesize that modality influences learners' sense of presence (H1), questioning behavior (H2), and learning outcomes (H5). We further propose that presence mediates these relationships (H3), while learner characteristics moderate them (H4). The specific hypotheses are detailed below.

\begin{enumerate}
  \item[\textbf{H1.}] \textbf{Direct Effect of Presentation Modality on Presence}
  \begin{enumerate}
    \item[\textbf{H1-1.}] The AR-IVA condition will yield a significantly higher level of \textbf{spatial presence} than the Video-IVA condition.
    \item[\textbf{H1-2.}] The AR-IVA condition will yield a significantly higher level of \textbf{social presence} than the Video-IVA condition.
  \end{enumerate}
\end{enumerate}

\noindent

Our first hypothesis investigates the impact of presentation modality on learners’ sense of presence. Drawing on the CAMIL model~\cite{makransky2021cognitive}, we focus on spatial and social presence as the primary dimensions of interest, as they most directly reflect the interactive characteristics of IVAs. In the AR-IVA condition, virtual learning objects are integrated into the learner’s physical environment, and the virtual instructor is presented as an embodied interaction partner. Because such spatial registration and embodied social cues are expected to induce a more immersive experience than video-based presentations, we hypothesize that spatial presence (H1-1) and social presence (H1-2) will be significantly higher in the AR-IVA condition than in the Video-IVA condition.


\begin{enumerate}
  \item[\textbf{H2.}] \textbf{Direct Effects of Presentation Modality on Questioning Behavior}

  \begin{enumerate}
    \item[\textbf{H2-1.}] \textbf{(When)} The AR-IVA condition will elicit significantly more questions during the \textbf{action stages} of the instruction compared to the Video-IVA condition.

    \item[\textbf{H2-2.}] \textbf{(What)}
    \begin{enumerate}
      \item[\textbf{H2-2a.}] \textbf{Topic:} The AR-IVA and Video-IVA conditions will elicit significantly different distributions of \textbf{question topics}.
      \item[\textbf{H2-2b.}] \textbf{Type:} The AR-IVA condition will have a significantly higher proportion of \textbf{procedural questions} and a lower proportion of \textbf{exploratory questions} compared to the Video-IVA condition.
      \item[\textbf{H2-2c.}] \textbf{Form:} The AR-IVA and Video-IVA conditions will elicit significantly different distributions of grammatical forms, with AR-IVA learners more frequently asking \textbf{Where}, \textbf{When}, \textbf{How}, and \textbf{HowMany} questions, whereas Video-IVA learners more frequently ask \textbf{Why} questions.
    \end{enumerate}

    \item[\textbf{H2-3.}] \textbf{(Why)} The AR-IVA and Video-IVA conditions will elicit significantly different patterns of learner \textbf{question intention}, reflecting modality-specific information-seeking strategies.

    \item[\textbf{H2-4.}] \textbf{(How)} The AR-IVA condition will show a significantly higher \textbf{volume (H2-4a)} of questions and higher linguistic \textbf{complexity (H2-4b)} compared to the Video-IVA condition.
  \end{enumerate}
    \end{enumerate}

Our second hypothesis examines the direct effects of presentation modality on learners’ questioning behavior. In the AR-IVA condition, learners adopt the perspective of an embodied actor~\cite{genay2021being,wolf2022exploring} by integrating instructional guidance within their physical space and establishing situational grounding between virtual instructions and potential real-world actions~\cite{radu2014augmented,henderson2010exploring,woodward2022analytic}. This situated perspective is expected to shift learners’ attention toward concrete action execution and procedural details.

Accordingly, we predict that in the AR-IVA condition, learners will ask questions more frequently during stages involving physical actions (H2-1) and focus more on procedural topics, types, and forms, characterized by increased use of process-oriented interrogatives such as \textit{where, when, how,} and \textit{how many} (H2-2)~\cite{anthony2004student}. We further hypothesize that learners’ intention will center on clarifying and verifying physical steps (H2-3). Finally, because mapping instructions onto potential actions requires precise procedural specification, we expect an overall increase in question volume and linguistic complexity in the AR-IVA condition (H2-4).

In contrast, in the Video-IVA condition, learners are expected to adopt an observer perspective, viewing instructions indirectly through a flat display~\cite{makransky2021cognitive,chi2014icap}. This video-based presentation format provides fewer spatial and situational cues, thereby limiting opportunities for situational grounding. As a result, learners’ attention is expected to shift away from action execution toward conceptual understanding. Accordingly, we anticipate a stronger emphasis on exploratory inquiry, particularly \textit{why} questions to seek conceptual or causal explanations (H2-2)~\cite{anthony2004student}. Moreover, because there is less need to specify concrete spatial or procedural details, we expect fewer questions overall and lower linguistic complexity than in the AR-IVA condition (H2-4).


\begin{enumerate}
  \item[\textbf{H3.}] \textbf{Mediation Effect of Presence}
  \begin{enumerate}
    \item[\textbf{H3-1.}] \textbf{(When)} Presentation modality will significantly and indirectly affect the \textbf{timing of questions} across instructional stages through the mediation of presence.
    
    \item[\textbf{H3-2.}] \textbf{(What)} Presentation modality will significantly and indirectly affect the \textbf{topic (H3-2a)}, \textbf{type (H3-2b)}, and \textbf{form (H3-2c)} of questioning through the mediation of presence.
    
    \item[\textbf{H3-3.}] \textbf{(Why)} Presentation modality will significantly and indirectly affect the \textbf{distribution of intention} through the mediation of presence.
    
    \item[\textbf{H3-4.}] \textbf{(How)} Presentation modality will significantly and indirectly affect the \textbf{volume (H3-4a)} and \textbf{complexity (H3-4b)} through the mediation of presence.
  \end{enumerate}
\end{enumerate}
\noindent

Our third hypothesis posits presence as a key psychological mechanism linking presentation modality to learners’ questioning behavior. Prior research has shown that presence mediates the impact of immersive features on active engagement and behavioral outcomes~\cite{lee2011effects, parong2020mediating}. Building on the premise that AR-IVA enhances presence (H1) and alters questioning strategies (H2), we hypothesize that spatial and social presence mediate the effects of presentation modality on learners’ questioning behavior, including timing (H3-1), content focus (H3-2), intention (H3-3), and volume and complexity (H3-4).


\begin{enumerate}
  \item[\textbf{H4.}] \textbf{Moderating Effect of Learning Motivation and Strategies}
  \begin{enumerate}
    \item[\textbf{H4-1.}] Learning motivation will significantly moderate the effect of presentation modality on questioning behavior, including 
    \textbf{When: the timing of questions (H4-1a)}, 
    \textbf{What: the content of questions (H4-1b)}, 
    \textbf{Why: the intention of questions (H4-1c)}, and 
    \textbf{How: the volume and complexity of questions (H4-1d)}.

    \item[\textbf{H4-2.}] Learning strategies will significantly moderate the effect of presentation modality on questioning behavior, including 
    \textbf{When: the timing of questions (H4-2a)}, 
    \textbf{What: the content of questions (H4-2b)}, 
    \textbf{Why: the intentions of questions (H4-2c)}, and 
    \textbf{How: the volume and complexity of questions (H4-2d)}.
  \end{enumerate}
\end{enumerate}

\noindent
Our fourth hypothesis proposes that learning motivation and strategies moderate the effects of presentation modality on questioning behavior. Prior SRL research suggests that questioning and help-seeking are not merely reactive to the learning environment but are strategically mediated by learners’ motivational beliefs and regulatory strategies \cite{newman2008adaptive, zimmerman2008investigating, xu2022understanding}. However, given the multifaceted nature of these individual differences, formulating specific directional hypotheses for each distinct subscale is practically challenging and arguably premature.

Accordingly, we adopt a broad moderation hypothesis. We expect learner motivation (H4-1) and learning strategies (H4-2) to moderate the effects of presentation modality on questioning behavior across the dimensions of timing, content focus, intention, volume, and complexity. Any significant interaction between presentation modality and a sub-component of these constructs will be interpreted as evidence of moderation, with specific subscale effects further examined through exploratory follow-up analyses.

\begin{enumerate}
  \item[\textbf{H5.}] \textbf{Direct Effect of Modality on Learning Outcomes} \\
  Learners in the AR-IVA condition will demonstrate significantly higher learning performance than those in the Video-IVA condition.
\end{enumerate}

Our fifth hypothesis predicts that learners in the AR-IVA condition will demonstrate superior learning performance compared to the Video-IVA condition. This expectation aligns with prior findings that the high immersion of AR environments enhances knowledge acquisition through embodied procedural learning~\cite{zhang2019applying,familoni2024augmented,wei2025towards}.

\subsection{System Setup}

\begin{figure}[t]
\centering
 \includegraphics[width=\linewidth]{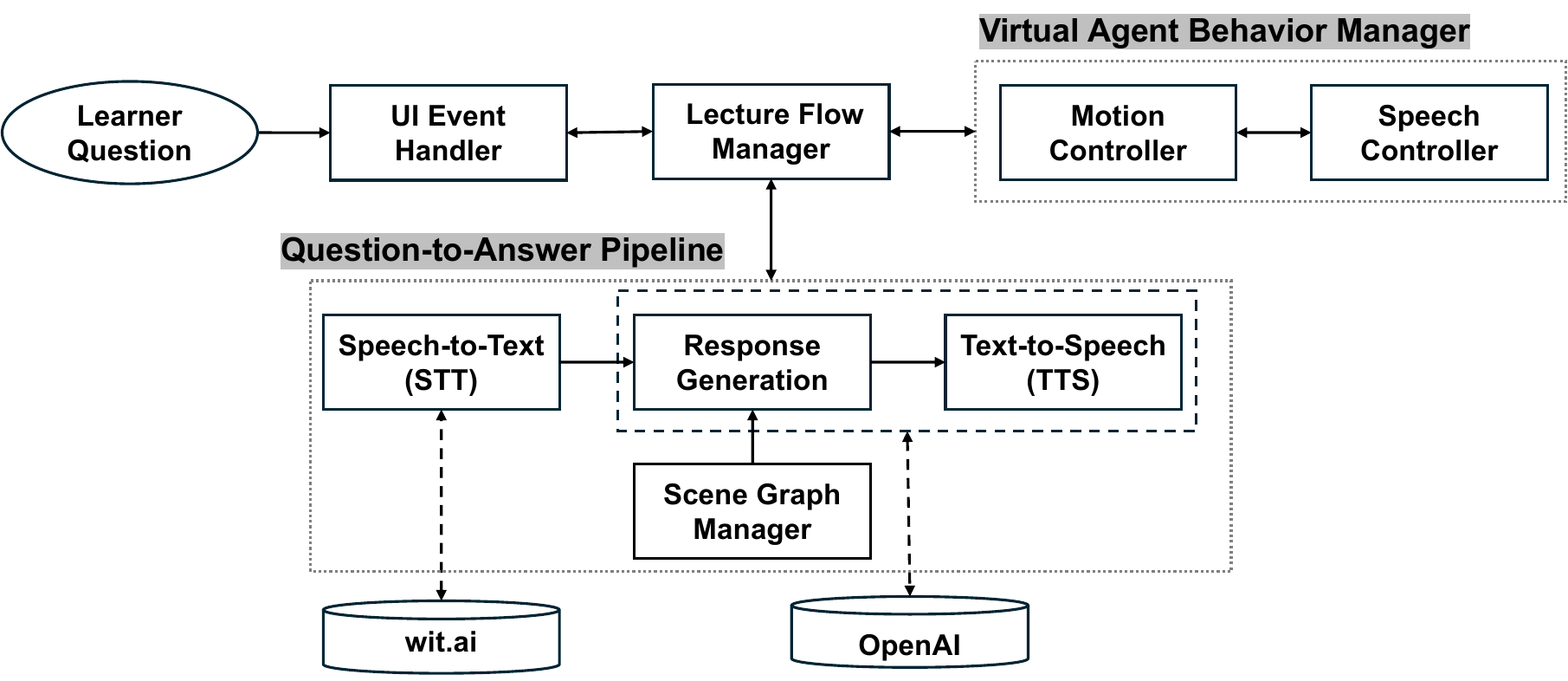}
    \caption{System architecture of the real-time Q\&A virtual agent system (Dashed boxes indicate grouped modules, and dashed lines indicate calls to external services).}
    \label{fig1}
    \Description{This figure shows the System architecture of the real-time Q\&A virtual agent system. Shaded boxes indicate grouped modules, and dashed lines indicate calls to external services. It begins with the Learner Question, where the user interacts with the Virtual Agent Behavior Manager. The Question to Answer Pipeline describes the Speech Flow generated by the Agent in response to the Learner's question. The external services and functions used in each module are explained.
}
\end{figure}

To empirically test our hypotheses, we first selected CPR as the instructional content. CPR training combines procedural actions with conceptual explanations for each step. Based on this structure, we considered that the instruction naturally elicits both procedural and conceptual questions. After determining the instructional content, we developed a Q\&A IVA system. The overall architecture of the system is shown in Figure~\ref{fig1}. The system runs within a Unity environment and consists of four main components: (1) Virtual Agent Behavior Manager, (2) UI Event Handler, (3) Lecture Flow Manager, and (4) Question-to-Answer Pipeline. The specific details are as follows.

The system delivers the CPR instruction using a fully rigged avatar with synchronized motion and narration. To enable this, instructor motions were captured from a 3-minute 14-second CPR tutorial using an OptiTrack marker-based suit and converted to FBX in Blender. The resulting animations were then applied to a Ready Player Me avatar via Unity’s Timeline.
Instructional audio was transcribed using Whisper and then resynthesized into sentence-level segments using OpenAI TTS API (\emph{echo} voice). Lip sync was implemented using BlendShape animation. The \emph{Virtual Agent Behavior Manager} coordinates the \emph{Motion Controller} and \emph{Speech Controller} modules to ensure that animation, narration, and pause/resume behavior remain aligned.

\begin{figure*}[t]
\centering
  \begin{subfigure}[t]{0.24\linewidth}
    \centering
    \includegraphics[width=\linewidth]{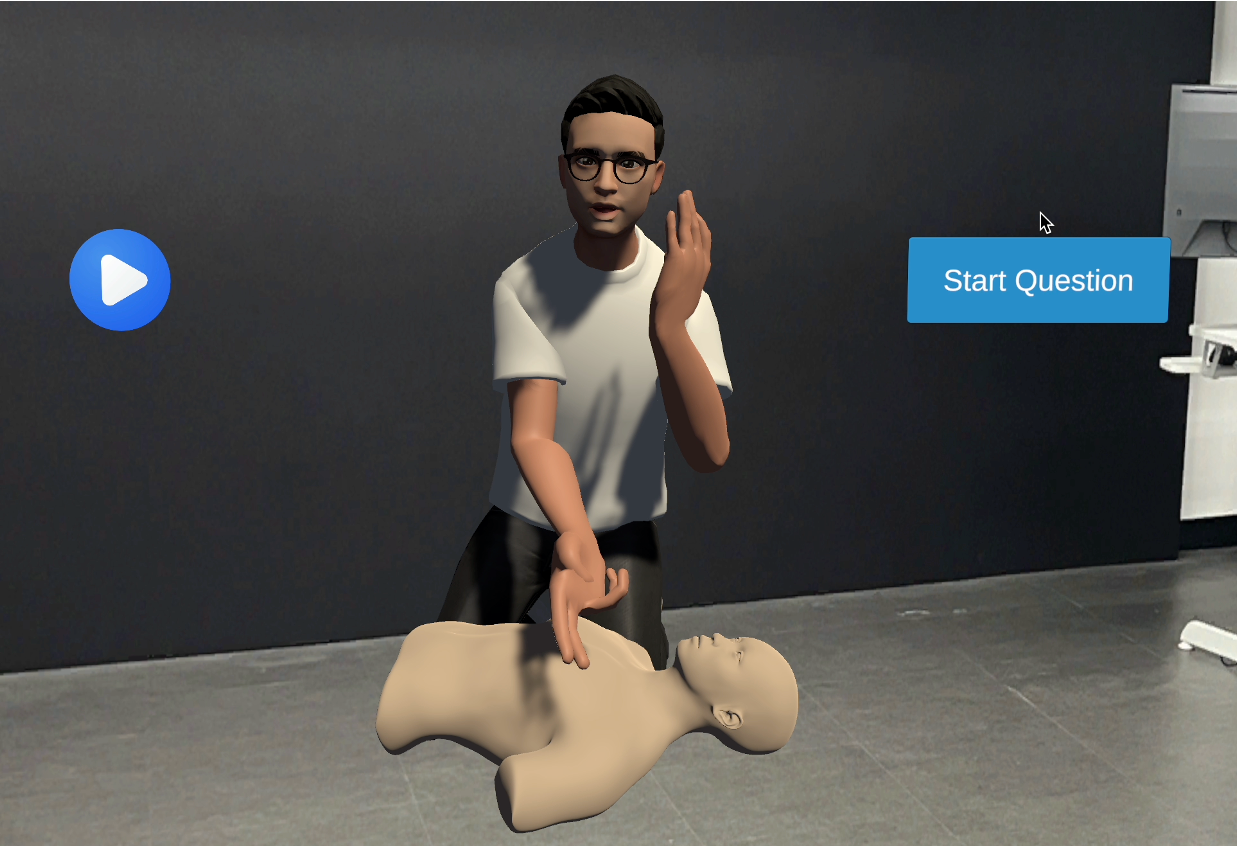}
    \caption{Instruction in Progress}
    \label{fig:instructing}
  \end{subfigure}
  \begin{subfigure}[t]{0.24\linewidth}
    \centering
    \includegraphics[width=\linewidth]{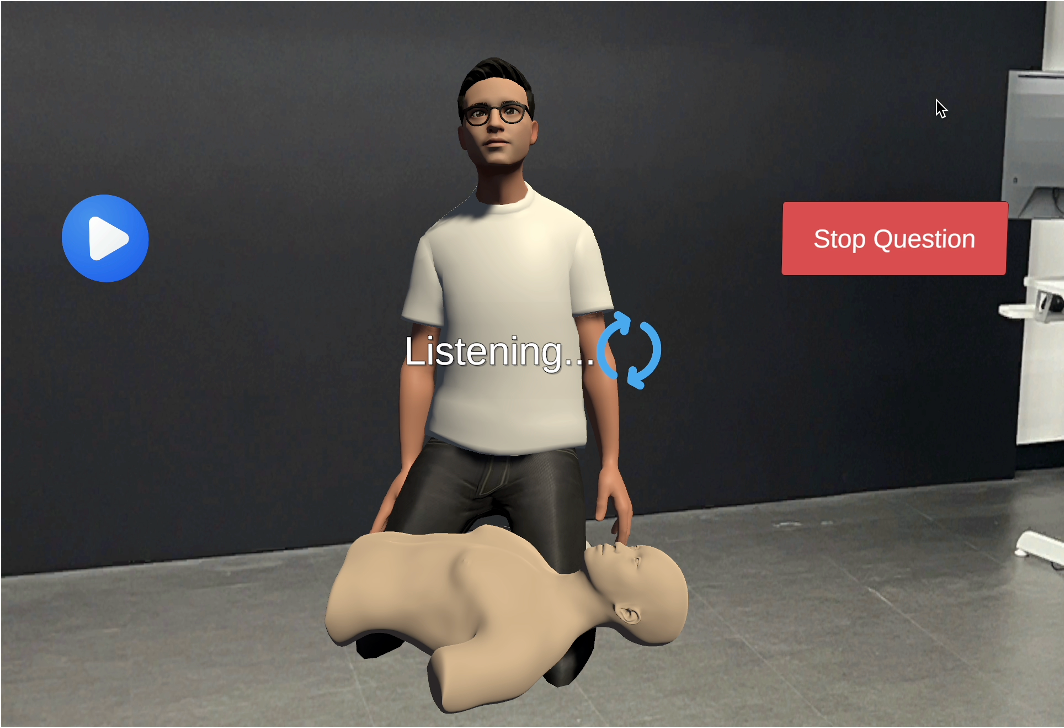}
    \caption{Listening}
    \label{fig:listening}
  \end{subfigure}
  \begin{subfigure}[t]{0.24\linewidth}
    \centering
    \includegraphics[width=\linewidth]{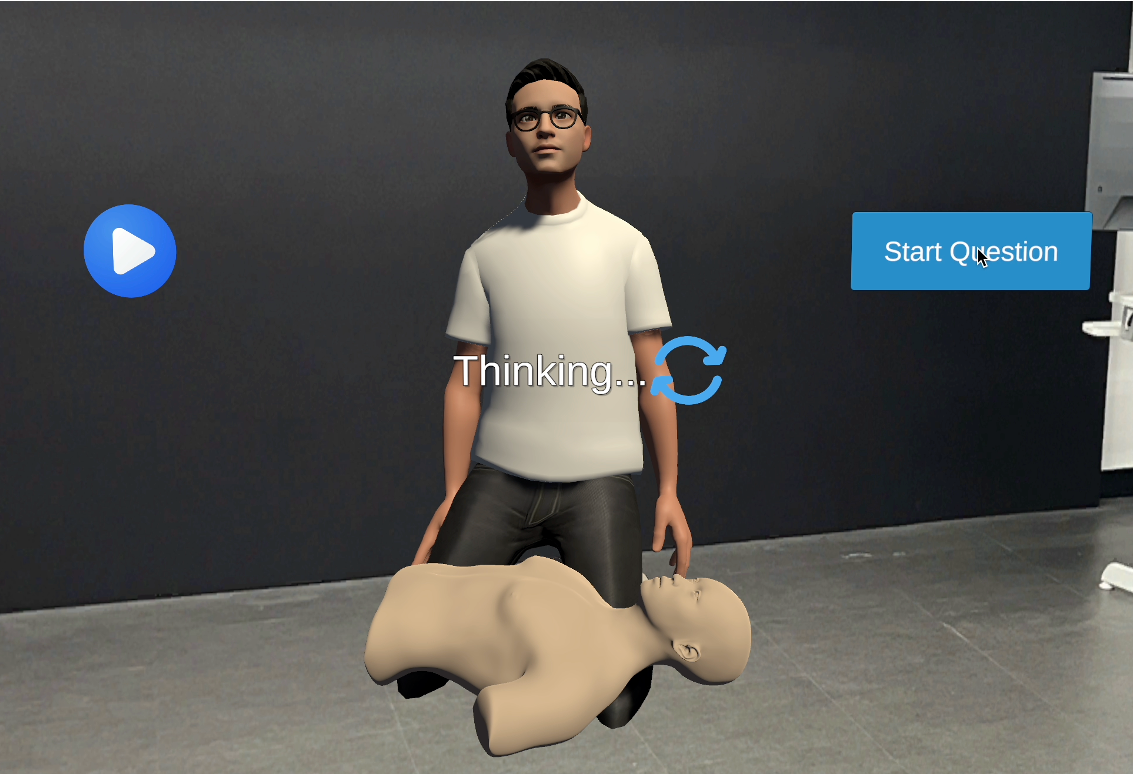}
    \caption{Thinking}
    \label{fig:thinking}
  \end{subfigure}
   \begin{subfigure}[t]{0.24\linewidth}
    \centering
    \includegraphics[width=\linewidth]{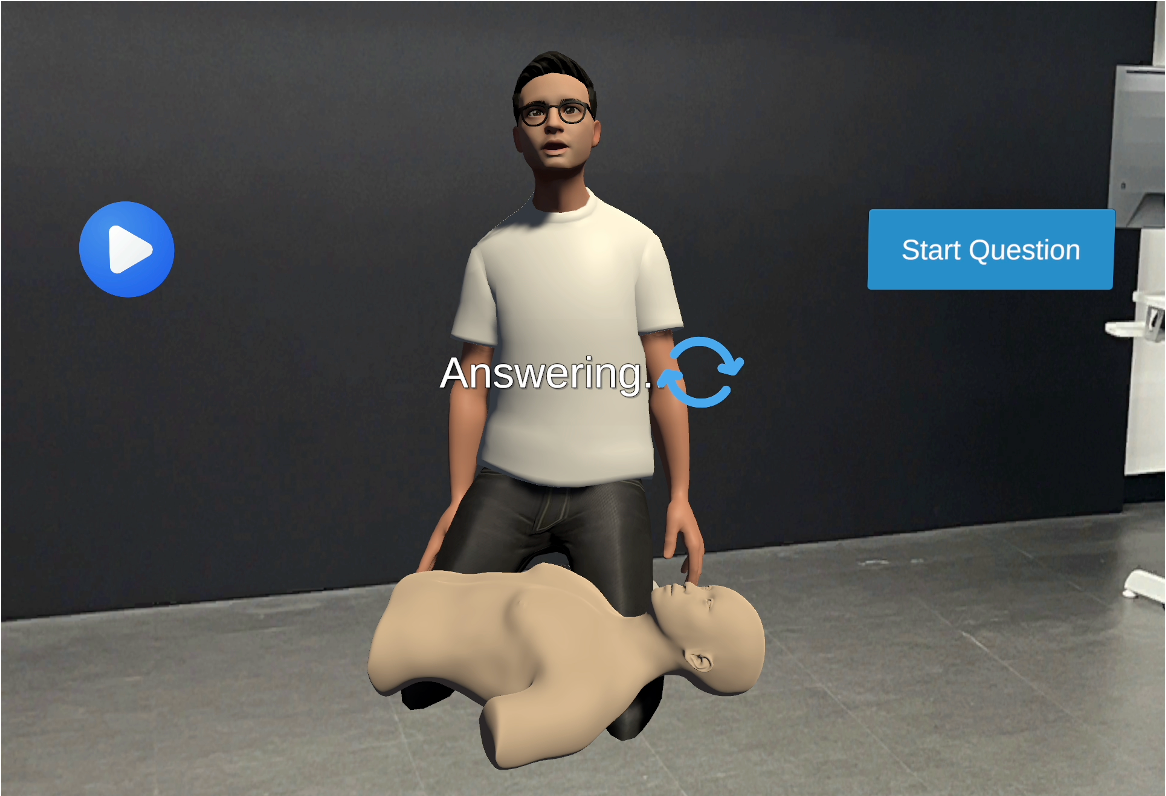}
    \caption{Answering}
    \label{fig:answering}
  \end{subfigure}

\caption{User interface feedback during interaction with the avatar-based IVA.}
\label{fig:feedback}
\Description{
The figure illustrates the sequential states of user interface feedback during interaction with the virtual agent.
The figure (a) depicts the ``Instruction in Progress'' state, where the agent delivers pre-defined learning content before any user inquiry.
The figure (b) shows the ``Listening'' state, which activates when the learner presses the ``Start Question'' button to input a query.
The figure (c) displays the ``Thinking'' state as the system processes the question and generates a response after receiving the learner's question.
The figure (d) illustrates the ``Answering'' state, where the agent provides a verbal explanation to the learner. This is provided as a speech-based explanation, and the image depicts that moment.
}
\end{figure*}

Interaction is primarily managed by the \emph{UI Event Handler} module, which captures all learner-initiated inputs—such as the Start/Stop Question toggle (Figure~\ref{fig:feedback})—and updates the interface accordingly. This module was adapted to different platforms depending on the experimental condition: a Meta Quest 3 for AR-IVA and a laptop for Video-IVA. Input methods were correspondingly tailored. Learners in the AR-IVA condition used the controller’s trigger button, whereas those in the Video-IVA condition performed the same actions via mouse clicks or trackpad input. In both conditions, learners’ spoken inquiries were recorded through each device’s built-in microphone.


All interaction events are forwarded to the \emph{Lecture Flow Manager} module, which coordinates transitions between instruction and Q\&A sessions. When a question is initiated, the manager pauses the instructional timeline and activates the avatar’s listening posture. After the IVA generates and delivers its spoken response, the manager resumes the lecture from the paused frame and restores the original animation layers. It also manages the UI feedback states (Listening → Thinking → Answering) (Figure~\ref{fig:feedback}) and logs all interaction timestamps.

The \emph{Question-to-Answer Pipeline} asynchronously converts spoken questions into a synthesized verbal response, operating independently of the lecture playback. First, the \emph{Speech-to-Text (STT)} module captures and transcribes the learner’s speech using the Meta Voice SDK (Wit.ai). When the question mode ends, the final transcript is forwarded to the \emph{Response Generation} module.
In this module, the system consults the instructional scene graph via the \emph{Scene Graph Manager}. The graph encodes the action–object relations, temporal stages, and instructor utterances extracted from the captured CPR demonstration. The learner’s transcribed question is combined with this contextual information and submitted to the OpenAI Chat Completions API (gpt-4o-mini). If the relevant scene-graph context is unavailable, the system automatically falls back to domain-general responses to maintain conversational continuity.
Finally, the \emph{Text-to-Speech (TTS)} module synthesizes the generated answer using the \emph{echo} voice, the same one used to resynthesize the instructional narration. The avatar plays the generated audio clip with synchronized lip sync.

\subsection{Procedure}

\begin{figure*}[t]
\centering
  \begin{subfigure}[t]{0.32\linewidth}
    \centering
    \includegraphics[width=\linewidth]{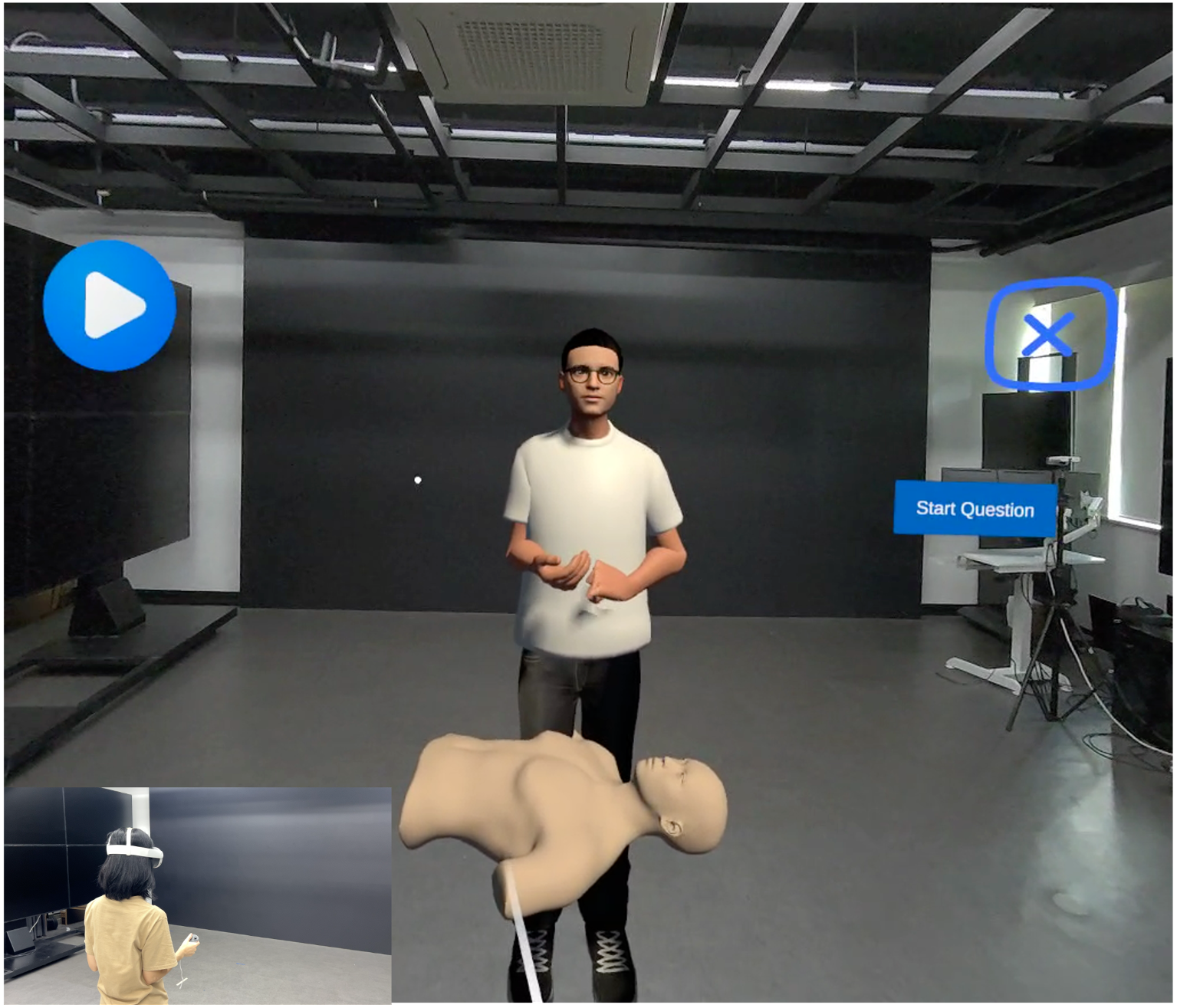}
    \caption{AR-IVA condition (inset: participant wearing a Meta Quest 3 headset and using a controller)}
    \label{fig:AR-IVA}
  \end{subfigure}
  \begin{subfigure}[t]{0.32\linewidth}
    \centering
    \includegraphics[width=\linewidth]{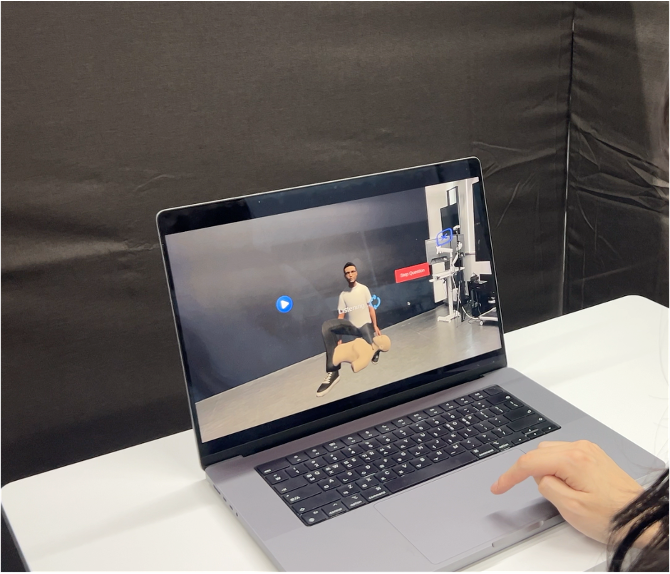}
    \caption{Video-IVA condition}
    \label{fig:Video-IVA}
  \end{subfigure}
  \begin{subfigure}[t]{0.32\linewidth}
    \centering
    \includegraphics[width=\linewidth]{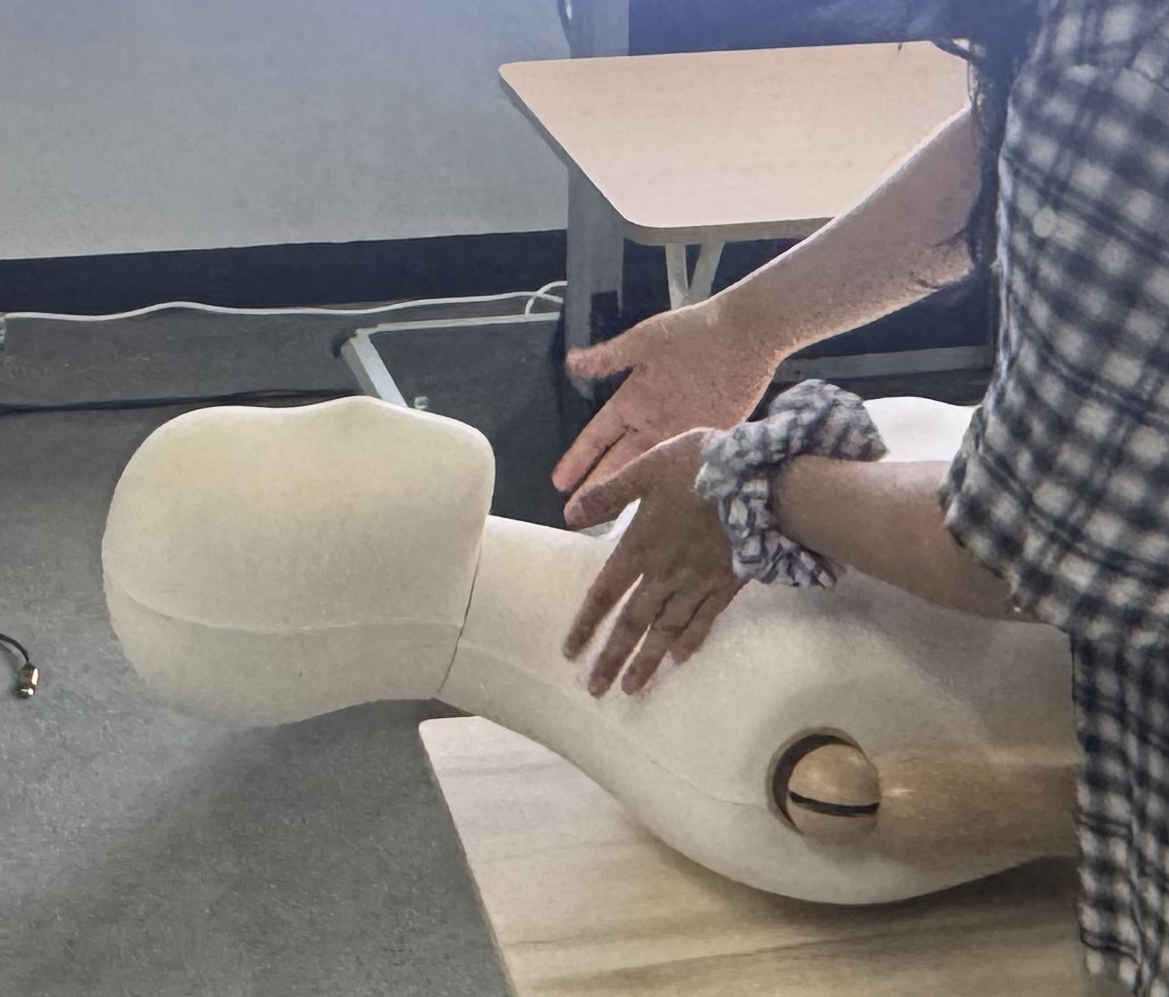}
    \caption{CPR practical performance evaluation using a manikin}
    \label{fig:practical}
   \end{subfigure}
\caption{Experimental setup.}
\label{fig:condition}
\Description{This figure details the physical and virtual setups for the two experimental conditions used to evaluate learner questioning behavior.
Figure (a) represents the AR-IVA (Augmented Reality Intelligent Virtual Agent) condition. In this setup, the learner wears a Meta Quest 3 headset enabling full-color passthrough mode. This allows the virtual instructor and the 3D overlay to be spatially anchored to the physical CPR manikin, creating a Mixed Reality (MR) experience. The image captures the learner's first-person perspective, demonstrating how digital information is superimposed onto the real-world training environment. The inset demonstrates the participant using a controller to point at and click the 'Start Question' button to engage with the system.
Figure (b) represents the Video-IVA condition, which serves as the control group. This setup utilizes a standard laptop display to present the identical instructional content and virtual agent avatar as a 2D video. Interaction in this condition is mediated through a standard computer mouse. In this condition, the participant interacts using a standard computer mouse to navigate the cursor and click the on-screen 'Start Question' button.
Figure (c)  depicts the physical setup used for the Practical Performance Evaluation, which serves as a primary dependent variable for assessing Learning Outcomes. Unlike the instructional phase, where virtual agents provided guidance, this evaluation takes place in a real-world environment without any virtual assistance. The image illustrates a participant checking the responsiveness of a physical CPR manikin. This setup ensures that the learning results are measured based on the participant's actual execution of the standard CPR protocol after the training session.
}
\end{figure*}

To examine learner questioning behavior with the developed system, we conducted a comparative user study.
The study was approved by the Institutional Review Board (KAISTIRB-2025-25), and all participants provided written informed consent. The experiment began with an introduction to the overall procedure and study purpose. To reduce ambiguity in asking questions, participants were shown examples of two question types (procedural or exploratory). They were also informed that they could ask questions freely during and after it had concluded. Finally, they were notified that a knowledge quiz and a practical performance evaluation would follow the learning phase.

Participants were randomly assigned to one of two conditions: AR-IVA or Video-IVA. Both conditions featured identical instructional content, system logic, and Q\&A interaction functionality. The main difference between the two conditions was how the virtual CPR instructor and the manikin were spatially represented, as shown in Figure~\ref{fig:condition}. Before the experiment, participants underwent a tutorial to familiarize themselves with the system, including launching the app, starting playback, and posing questions in their assigned experimental condition. Participants’ experiences in each condition were as follows.

\textbf{AR-IVA Condition:} Participants wore a Meta Quest 3 headset and launched the Unity-based IVA application by pressing the controller’s trigger button on the home screen. Once the app started, the CPR instructor avatar and manikin appeared as spatially anchored 3D overlays in the participant’s real environment (Figure.~\ref{fig:AR-IVA}). Using the controller’s trigger button, participants pressed the playback button on the left to start the instruction and used the question button on the right to initiate and end a question at any time. After completing the lecture and all questions, they pressed the exit button in the upper-right corner to finish the session.

\textbf{Video-IVA Condition:} Participants ran the same application on a 15-inch laptop, where the identical content was rendered as a 2D view (Figure.~\ref{fig:Video-IVA}). To maintain visual consistency, the background displayed a static image of the physical environment of the AR-IVA condition. Instead of spatial interaction, participants used the trackpad to click the on-screen play, question, and exit buttons. This condition was designed to simulate a typical online video learning experience.

In both conditions, the user interface provided real-time visual feedback to indicate the current state of interaction. When the learner activated the `Start Question' button (blue) (Figure~\ref{fig:instructing}), the system began recording and displayed a ``Listening...'' status panel (Figure~\ref{fig:listening}). When the learner pressed the `Stop Question' button (red), the panel transitioned to ``Thinking...'' while the system processed the input (Figure~\ref{fig:thinking}). During answer playback, the interface showed ``Answering...'' while the avatar provided the synthesized verbal response synchronized with lip-sync (Figure~\ref{fig:answering}). Upon completion of the response, the panel disappeared, and the instructional timeline automatically resumed after a 2-second delay.

Following the learning session, participants first completed a knowledge quiz and then performed a practical evaluation using a CPR manikin, as shown in Figure~\ref{fig:practical}. During this evaluation, participants were required to verbally state or physically execute each step. If a participant expressed uncertainty about a specific step, it was bypassed to prevent undue stress. Subsequently, participants reviewed the saved text logs of their questions and assigned intention labels by selecting the most appropriate category from a predefined taxonomy. 
Finally, participants completed a set of questionnaires administered via Google Forms in the following order: (1) demographic information, (2) System Usability Scale (SUS), (3)  NASA Task Load Index (NASA-TLX), (4) Motivated Strategies for Learning Questionnaire (MSLQ), (5) Networked Minds Measure of Social Presence (NM) and the Temple Presence Inventory (TPI), and (6) MEC Spatial Presence Questionnaire. The study concluded with a semi-structured interview, which explored participants’ rationale behind their questioning and their perceptions of learning. Detailed descriptions of all measurement instruments and evaluation metrics are provided in Section~\ref{Measurements}.


\subsection{Measurements}
\label{Measurements}
To comprehensively evaluate the impact of presentation modality, we employed a multi-dimensional measurement strategy. This section is organized into two subsections. First, we detail how questioning behavior is quantified using our 5W1H-based framework. Second, we describe the assessment of psychological constructs and learning outcomes. The specific details are described below. All questionnaire items and implementation details are provided in the supplementary materials.

\subsubsection{Operationalization of the 5W1H-based Framework.}
\label{5W1H_Dimensions}
To systematically analyze learner questioning, we operationalized the proposed 5W1H-based framework into concrete measures, leveraging both recorded user queries and participant-reported data.

\textbf{Who. }
To account for learner characteristics, we utilized the Motivated Strategies for Learning Questionnaire (MSLQ)~\cite{duncan2005making}. The motivation section included all six original subscales (31 items), assessing value beliefs (Intrinsic/Extrinsic Goal Orientation, Task Value), expectancy (Control of Learning Beliefs, Self-Efficacy), and affect (Task Anxiety). These scales capture the motivational drivers underlying learners’ questioning behavior. For learning strategies, we selected five subscales (32 items) that are most relevant to agent-based, modality-dependent learning, focusing on cognitive skills (Critical Thinking, Metacognitive Self-Regulation) and resource management tactics (Time and Study Environment Management, Effort Regulation, Help Seeking). These constructs were selected to examine how learners’ self-regulatory profiles moderate their questioning behavior.

\textbf{When. }
To analyze the temporal characteristics of questioning, we used two primary metrics: normalized time position and instructional stage. Each question was assigned a timestamp relative to the session start. This timestamp was then normalized to a [0, 1] scale to enable comparisons across sessions of different lengths. The instructional timeline was segmented into four stages—\emph{Introduction}, \emph{Explanation}, \emph{Action}, and \emph{Ending}. The \emph{Introduction} and \emph{Ending} stages were manually annotated based on the instructor’s opening and closing remarks. The \emph{Action} stage was automatically identified using action nodes extracted from the instructional scene graph. The remaining portions of the lecture were labeled as the \emph{Explanation} stage. Questions were then mapped to these stages based on their timestamps and manually verified for accuracy.

\textbf{What. }
To analyze the semantic content of learner questions, we characterize each query along three components: topic, type, and form. Before analysis, all queries undergo pre-processing: tokenization, lowercasing, normalization of whitespace and repeated punctuation, and standardization of orthographic variants/synonyms.
For the implementation details and hyperparameters (including confidence thresholds of each analysis), and prompting methods of LLM topic modeling (topic) and LLM disambiguation (type and form), please consult the supplementary materials.

\textbf{(1) Topic Extraction. }
Since the queries in our case are short and domain-specific, conventional probabilistic or clustering-based approaches tended to collapse into a single dominant cluster (\textit{e.g.}, \texttt{CPR}, \texttt{chest}). Thus, to extract topics from queries, we employ a ``discover–assign–consolidate'' pipeline, based on LLM-assisted topic modeling and labeling~\cite{wang2023llmtopicmodeling, topicgpt}.

(i) \textit{Discover. } 
For each query $s$, we extracted a small set of keyphrases $K(s)$ through standard text preprocessing (n-gram and TF–IDF–based keyword extraction).
We then instruct an LLM to induce a concise catalog $\mathcal{T}$ of $K$ domain topics,
\begin{equation}
\mathcal{T}=\{(t_i,\ \ell_i,\ \mathrm{desc}_i,\ \mathrm{keys}_i)\}_{i=1}^{K},
\end{equation}
where each topic $t_i$ is specified by a short label ($\ell_i$), a one-sentence definition ($\mathrm{desc}_i$), and a small set of canonical keyphrases ($\mathrm{keys}_i$). This catalog $\mathcal{T}$ served as the semantic foundation for subsequent topic assignment.

(ii) \textit{Assign. } 
Given the predefined topic catalog $\mathcal{T}$, we computed a confidence score for each topic $t_i$ by providing the LLM with the raw text of each query together with its extracted keyphrases $K(s)$. For each query $s$, we then selected up to $M$ topics with the highest scores, provided that they exceeded a minimum confidence threshold. To prevent cases where the LLM abstained from assigning any topic, we applied a lightweight fallback strategy based on keyword overlap between $K(s)$ and the representative keyphrases in $\mathrm{keys}_i$. If no match was found, the query was assigned to a general ``Other/General'' category.

(iii) \textit{Consolidate. } 
Finally, we consolidated the catalog by merging very small topics into the most compatible larger ones based on keyword similarity and LLM tie–breaks and remapped affected assignments (e.g., combining ``AED location'' and ``attachment'' into ``AED Placement \& Usage'').  
This removes spurious fragments while preserving interpretable topic boundaries and improving coverage.

\textbf{(2) Type Dichotomization. } 
To determine whether learners' inquiries were \textit{Exploratory} or \textit{Procedural}, we adopted a hybrid cascade approach combining rule-based scoring with LLM disambiguation, following prior work~\cite{banerjee2012questionclassification, madabushi2016questionclassification, wang2023llmselfconsistency}. Concretely, we accumulate pattern-hit counts into two rule scores ($E(s)$: \emph{Exploratory}, $P(s)$: \emph{Procedural}) for query $s$:
\begin{equation}
\begin{split}
E(s) &= \textstyle\sum_{l\in\mathcal{L}_E}\!\!\mathrm{hits}_l(s) + \delta_{\textsf{why}}(s), \\
P(s) &= \textstyle\sum_{l\in\mathcal{L}_P}\!\!\mathrm{hits}_l(s) + \delta_{\textsf{pp}}(s),
\end{split}
\end{equation}

where $\mathrm{hits}_l(s)$ denotes the number of times the lexicon entry or regex pattern  $l$ appears in the tokens of $s$.
$\delta_{\textsf{why}}(s)$ and $\delta_{\textsf{pp}}(s)$ serve as semantic weighting terms to handle short, ambiguous queries. $\delta_{\textsf{why}}(s)$ boosts the exploratory score for ``Why'' questions (signaling conceptual intent). By contrast,  $\delta_{\textsf{pp}}(s)$ boosts the procedural score for ``Where'', ``When'', and ``How many'' questions (signaling factual or step-oriented intent). 
These adjustments were applied to stabilize the classification, particularly for short queries with limited lexical cues.

$\mathcal{L}_E$ and $\mathcal{L}_P$ each combine lexicon entries and regex-style patterns. 
For $\mathcal{L}_E$, exploratory cues include words such as \textit{why}, \textit{because}, \textit{reason}, 
\textit{purpose}, and \textit{mechanism}. 
It also contains pattern templates that capture mechanism-oriented questions (e.g., questions of the form ``How does X work/help/affect Y?'' or ``What causes/leads to/results in X?''). 

In contrast, $\mathcal{L}_P$ captures procedural cues, including action-guidance terms such as \textit{how to} and \textit{should}, as well as locational and quantitative interrogatives such as \textit{where}, \textit{when}, and \textit{how many}. 
It also includes procedure-related expressions such as \textit{steps}, \textit{rate}, and \textit{position}. 
In addition, $\mathcal{L}_P$ includes numeric or measurement-oriented regex templates (e.g., expressions referring to rates such as ``X per minute'').

The confidence of the rule-based classification was computed by mapping the margin between the exploratory and procedural scores to a [0, 1] range via a sigmoid function. When this confidence fell below a predefined threshold, we performed LLM-based disambiguation using the raw text and the rule-based outputs (initial label and scores). The final label was determined using a confidence-triggered cascade. If the rule-based confidence exceeded the threshold, the rule-based label was retained; otherwise, it was replaced by the LLM-refined label. The final confidence value corresponds to that of the final selected classification stage.

\textbf{(3) Form Classification.} 
To identify the structural form of each query, we follow standard typologies of interrogatives and prior work on question classification~\cite{Karttunen1977, questionclassifier2002, wang2023llmselfconsistency}.
We adopt a hybrid classifier that first applies rule-based scoring and, when confidence is low, leverages an LLM with a self-consistency voting procedure for disambiguation.
We define a fixed class set $\mathcal{C} = \{\text{WH}, \text{YN}, \text{ALT}, \text{REQ}\}$, corresponding to information-seeking questions (WH: who/\allowbreak what/\allowbreak when/\allowbreak where/\allowbreak why/\allowbreak how), yes–no questions (YN), alternative questions (ALT), and requests or imperatives (REQ).

In the rule-based stage, we apply pattern matching using surface lexical cues such as WH-words (e.g., ``when,'' ``why''), clause-initial interrogative auxiliaries (e.g., ``is,'' ``do,'' ``can''), alternative markers (``or''), and request markers (e.g., ``please,'' ``could you''). 
For each class, the number of matching tokens is counted and used to compute class-specific scores.
They are then normalized with a softmax function to produce class probabilities. 
The highest probability is taken as the rule-based confidence.

When the rule-based confidence falls below a predefined threshold, we leverage an LLM. 
The model is provided with the raw query along with its immediate context (the preceding and following queries) and generates a probability distribution using self-consistency voting across multiple sampled outputs. 
After generating a set of candidate classes, the model selects the majority-voted class as the final prediction.
The mathematical formulation of the rule-based decision logic and the LLM disambiguation prompts are provided in the Supplementary Materials.

\textbf{How. }
We assess intensity and depth of a learner’s questioning behavior by measuring per-user query volume and per-query linguistic complexity.

\textbf{(1) Query Volume. }
We quantified per-user query volume, following prior work on utterance-based measures~\cite{CHEN201821, ghazarian2020}.
For each learner, we computed two indicators of query volume: \emph{query count}, defined as the total number of queries ($U$), and \emph{query length}, defined as the total token count ($T$) generated during the session.
To support comparison across modalities, both $U$ and $T$ were standardized by converting them into $z$-scores over the full participant pool.
These standardized values ($z_U$ and $z_T$) were then used for analysis.

\textbf{(2) Query Complexity. }
We quantified the structural and linguistic complexity of each query, following established methods~\cite{lu2010complexity}. For each query, we extract a fixed set of per-query features of complexity proxies (c1–c6): 
\textbf{(c1)} clauses per T-unit, 
\textbf{(c2)} dependent clause ratio, 
\textbf{(c3)} mean dependency distance, 
\textbf{(c4)} dependency tree depth, 
\textbf{(c5)} coordination/subordination and conditional marker rate (per token), and 
\textbf{(c6)} average token length as an inverse proxy for simplicity. We adopt six complexity proxies that capture complementary aspects of syntactic load. Clausal ratios (c1–c2) indicate the degree of subordination, while structural metrics (c3–c4) represent structural distance and hierarchical depth. Marker-based features (c5) capture coordination and conditional cues, and average token length (c6) serves as a lightweight proxy for lexical sophistication.

The six complexity features are computed using lightweight heuristics over tokenized queries. Clausal metrics (c1–c2) estimate clause counts from punctuation and subordination markers, with c2 representing the ratio of dependent clauses. Structural features (c3–c4) approximate dependency distance as the mean gap between repeated content words and measure hierarchical depth through parenthetical nesting. Marker-based features (c5) are extracted by matching predefined coordination/subordination lexicons against the text, normalized by token count. Lexical complexity (c6) is computed as the mean character length of all tokens.
Each feature was standardized using z-scores computed over the full participant pool. We then calculated a composite complexity score ($\mathrm{CC}$) for each query as a weighted sum of its standardized feature values. A higher $\mathrm{CC}$ score indicates greater structural and linguistic complexity. Finally, user-level complexity was obtained by averaging the $\mathrm{CC}$ scores of all queries from each user ($\overline{\mathrm{CC}}$).

\textbf{Why. }
To classify the underlying intention of learner questions, we adopted the motivation taxonomy proposed by Kim et al.~\cite{kim2024using}. Unlike traditional keyword-based frameworks, this taxonomy is optimized for dynamic, multi-turn dialogues where user intention evolves throughout the interaction. While automated methods such as LLM-based inference exist, they function primarily as post-hoc proxies. Therefore, to ensure high reliability and capture the ground truth of learners' specific goals, we employed a self-assessment method. Using this validated taxonomy, participants reviewed their interaction logs and annotated each question based on six intention types: (1) clarifying ambiguity, (2) seeking deeper understanding, (3) exploring related domains, (4) narrowing to specifics, (5) requesting alternative representations, and (6) verifying information.

\subsubsection{Dependent Variables}
To clarify the mechanisms underlying learner behaviors and to ensure the validity of the experimental findings, we assessed three categories of variables.

\textbf{Presence. } 
Spatial and social presence were measured both as an outcome of presentation modality and as a potential mediator shaping questioning behavior. Spatial presence was assessed using the MEC Spatial Presence Questionnaire~\cite{vorderer2004mec}, which includes 24 items (four per scale), rated on a five-point Likert scale. 
Social presence was assessed using items adapted from two validated instruments: the Networked Minds Measure of Social Presence (NM)~\cite{harms2004internal} and the Temple Presence Inventory (TPI)~\cite{lombard2009measuring}. Specifically, participants responded to 15 items from three NM subscales—Co-Presence (6 items), Attentional Allocation (self-focused; 3 items), and Perceived Message Understanding (6 items)—and 8 items from the TPI- Parasocial Interaction subscale (7 items) and Active Interpersonal (1 item). All social presence items were rated on a seven-point Likert scale.

\textbf{Learning Outcomes.}
To determine the instructional effectiveness of the system, we assessed both knowledge acquisition and CPR practical performance.
The knowledge quiz comprised eight items: four assessing procedural knowledge and four assessing conceptual understanding.
CPR practical performance was evaluated using a manikin-based, eighteen-item checklist aligned with the five CPR steps.
Each correctly performed action was scored as one point.

\textbf{Control Variables. }
Finally, system usability and perceived workload were measured to control for potential confounds related to system difficulty and interface demands. System usability was assessed with the System Usability Scale (SUS)~\cite{bangor2008empirical}, and perceived workload with the NASA Task Load Index (NASA-TLX)~\cite{hart2006nasa,hart1988development}. Full item lists for all instruments are provided in the supplementary materials.

\subsection{Data Analysis Strategy}

\textbf{Comparative Analysis. }
Presence (H1), question volume and linguistic complexity (H2-4), and learning performance (H5) were compared using independent-samples t-tests. Question timing across instructional stages (H2-1) was examined using Fisher's exact test with Monte Carlo simulation (5,000 replicates) to account for sparse data in certain instructional phases.
Question topic (H2-2a), form (H2-2c), and intention (H2-3) were analyzed as multiclass categorical variables using a two-stage approach. First, to assess between-condition differences, we conducted Fisher’s exact tests on the full contingency tables. When significant effects were observed, post-hoc pairwise Fisher's exact tests with false discovery rate (FDR) correction were performed (reported as adjusted p-values, $p_{adj}$). Subsequently, to examine distributional patterns within each condition, we conducted chi-square goodness-of-fit tests. Where significant deviations were detected, standardized residuals were used to identify over- or underrepresented categories, with $p$-values adjusted using the Holm–Bonferroni method (reported as $p_{adj}$).  Question type (H2-2b) was analyzed as a binary variable using a Fisher’s exact test.

\textbf{Mediation Analysis. }
Causal mediation analyses were performed to examine whether spatial or social presence mediates the effect of presentation modality on questioning behavior. For multiclass outcomes, including instructional stages (H3-1), topics (H3-2a) and forms (H3-2c), and intentions (H3-3), we employed a one-vs-rest decomposition strategy, utilizing logistic regression for the outcome models and linear regression for the mediator models. Average Causal Mediation Effects (ACME) were estimated via Quasi-Bayesian Monte Carlo simulation (1,000 draws), with FDR correction (reported as $p_{adj}$). For the binary question type (H3-2b), a single logistic mediation model was used, whereas continuous outcomes, including question volume and complexity (H3-4), were analyzed using linear regression. The \textit{Introduction} stage was excluded from temporal analyses due to the absence of learner interaction.

\textbf{Moderation Analysis.} 
To examine whether learner characteristics—specifically motivation (H4-1) and strategies (H4-2)—moderate the effects of presentation modality, we estimated regression models for each trait, incorporating condition, learner trait, and their interaction term as predictors. We employed multinomial logistic regression for multiclass outcomes, covering question timing (H4-1a, H4-2a), topics and forms (H4-1b, H4-2b), and intentions (H4-1c, H4-2c). Binomial logistic regression was used for the question type (H4-1b, H4-2b), and linear regression for question volume and complexity (H4-1d, H4-2d). 
Given the exploratory aim of identifying potential moderation patterns across multiple learner traits, we reported uncorrected $p$-values. 
These results were interpreted as indicative trends rather than confirmatory evidence.
To ensure model stability, we excluded outcome categories with fewer than five observations in any condition from the moderation analyses. 
This exclusion was applied to avoid estimation issues associated with sparse data.

\subsection{Participants} 
All participants (N = 40) were recruited through the university website. They were over 18 years of age and had normal or corrected-to-normal vision of 0.6 or higher. Their ages ranged from 22 to 36 years, with a mean of \textit{M} = 27.03 years (\textit{SD} = 3.37). Participants were randomly assigned to one of two presentation modalities: AR ($n$ = 20) or Video ($n$ = 20). In the AR-IVA condition, 20\% of participants reported being highly proficient in AR, 45\% indicated limited experience (used only a few times), and 35\% reported occasional use (approximately once every 2--5 months). In the Video-IVA condition, prior AR experience was not assessed, as the task did not involve AR-based interactions. Regarding prior CPR knowledge, participants self-reported their familiarity levels: 5\% indicated not know ($n$ = 2), 35\% reported slight knowledge ($n$ = 14), 40\% reported moderate knowledge ($n$ = 16), and 20\% reported good knowledge ($n$ = 8). At the session level, participants overall asked on average \textit{M} = 5.40 questions (\textit{SD} = 2.47) and spent \textit{M} = 7.93 minutes (\textit{SD} = 2.62). AR-IVA participants ($n$ = 20) asked more questions (\textit{M} = 6.60, \textit{SD} = 2.64) and spent longer (\textit{M} = 8.33 minutes, \textit{SD} = 2.95) than Video-IVA participants ($n$ = 20), who asked fewer questions (\textit{M} = 4.20, \textit{SD} = 1.58) and spent less time (\textit{M} = 7.53 minutes, \textit{SD} = 2.22). There were no significant differences between the AR-IVA and Video-IVA conditions on NASA-TLX and SUS.

\textit{Preregistration. }
The hypotheses and analysis plans were specified prior to data collection; however, the study was not formally preregistered.

\section{Results} 
In this section, we present the main quantitative findings, beginning with the direct effects of presentation modality on spatial and social presence and questioning behavior. Subsequently, we detail analyses of the underlying mechanisms, focusing on the mediating role of spatial and social presence and the moderating role of learners' motivation and strategies. Finally, we examine the effects on learning outcomes. To conclude, we provide qualitative insights from participant interviews to contextualize these quantitative results.

\subsection{Quantitative Analysis Results}

\subsubsection{Direct Effect of Presentation Modality on Presence}

\begin{figure}[t]
\centering
 \begin{subfigure}[t]{0.4\linewidth}
    \centering
    \includegraphics[width=\linewidth]{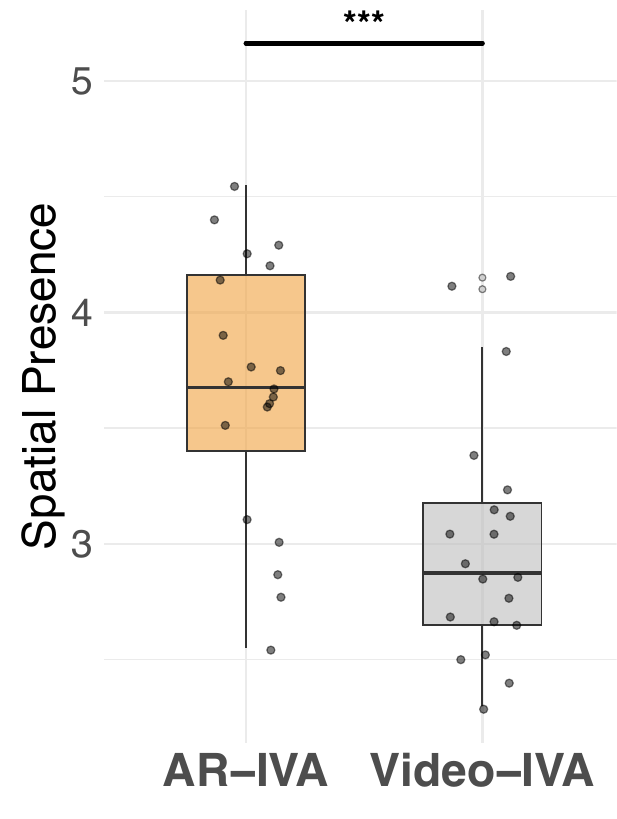}
    \caption{Spatial presence}
    \label{fig:qspatial_presence}
  \end{subfigure}
  \begin{subfigure}[t]{0.4\linewidth}
    \centering
    \includegraphics[width=\linewidth]{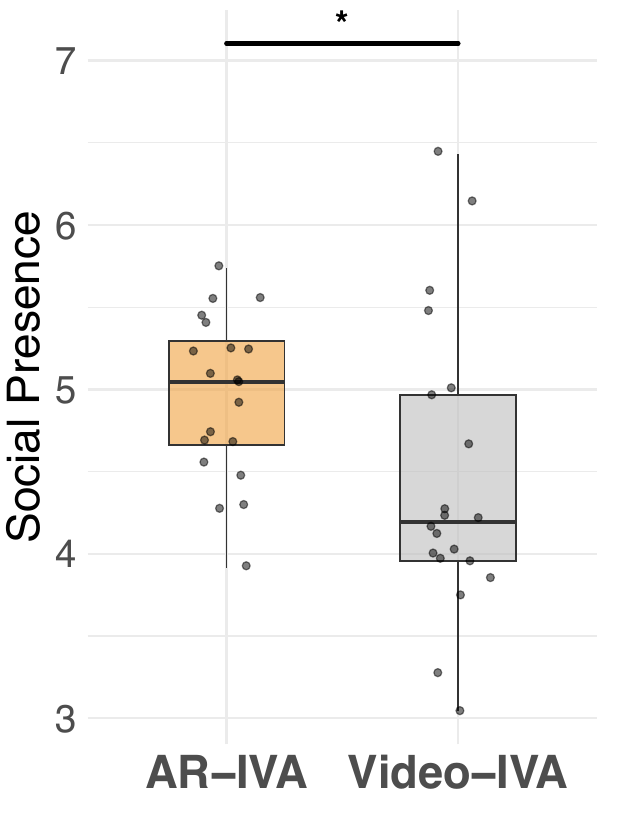}
    \caption{Social presence}
    \label{fig:social_presence}
  \end{subfigure}\caption{Effects of presentation modality on \emph{presence}. ($*:p<0.05$; $***:p<0.001$)}
\label{fig:h1h2-3a}
\Description{This figure presents the quantitative comparison of user experience metrics—specifically Spatial Presence and Social Presence—between the AR-IVA and Video-IVA conditions.
The data is visualized using box plots overlaid with scatter points to show both the central tendency and individual response distribution.
Spatial Presence (Left): The plot indicates that participants in the AR-IVA condition perceived a much stronger sense of being physically present in the learning environment compared to the Video-IVA group. The triple asterisks denote a highly significant statistical difference (p<.001), reflecting the substantial gap in mean scores shown in the graph.
Social Presence (Right): The results also show higher ratings for the AR-IVA condition regarding the sense of social connection with the virtual agent. The single asterisk indicates a statistically significant difference (p<.05). While the distribution in the Video-IVA condition is more spread out, the AR-IVA condition consistently elicited higher median scores.
Overall, these visual comparisons confirm that the AR-IVA (AR Modality) effectively enhances both the spatial realism and the social connection with the instructor relative to the 2D-IVA(Video Modality).
}
\end{figure}

The analysis revealed a significant effect of presentation modality on both spatial and social presence. As shown in Figure~\ref{fig:h1h2-3a}, participants in the AR-IVA condition reported higher spatial presence ($M = 3.66$, $SD = 0.57$) than those in the Video-IVA condition ($M = 3.01$, $SD = 0.53$), $t(37.78) = 3.75$, $p < .001$, supporting \textbf{H1-1}. Participants in the AR-IVA condition also reported higher social presence ($M = 4.96$, $SD = 0.49$) than those in the Video-IVA condition ($M = 4.46$, $SD = 0.89$), $t(29.60) = 2.20$, $p = .036$, supporting \textbf{H1-2}. These results demonstrate higher reported spatial and social presence in the AR-IVA condition relative to the Video-IVA condition.

\subsubsection{Direct Effects of Presentation Modality on Questioning Behavior}

\textbf{When. }
For \emph{question timing}, no significant association with condition was observed across instructional stages ($p = .320$), leading to a rejection of \textbf{H2-1}. 
This indicates that learners in both the AR-IVA and Video-IVA conditions tended to ask their questions at comparable points in the instructional sequence. It suggests that the timing of inquiries was shaped not by presentation modality but by other factors, such as the task structure and individual learner characteristics.


\begin{figure*}[t]
\centering
\includegraphics[width=.9\linewidth]{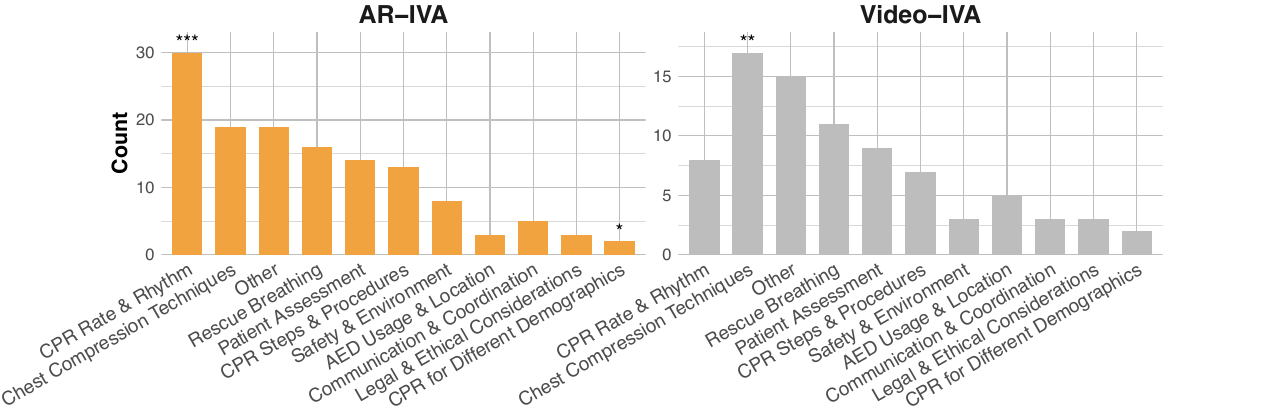}
\caption{Distribution of \emph{question topics} across AR-IVA and Video-IVA conditions. (*: $p_{adj} < 0.05$; **: $p_{adj} < 0.01$; ***: $p_{adj} < 0.001$)}
\label{fig:h2_topic}
\Description{This figure presents two bar charts comparing the distribution of question topics between the AR-IVA (left) and Video-IVA (right) conditions. The horizontal axis (x-axis) lists question topics such as "CPR Rate \& Rhythm," "Chest Compression Techniques," and "Rescue Breathing." The vertical axis (y-axis) represents the Count (frequency) of questions asked.
AR-IVA Condition (Left Chart): In the AR condition, the distribution is highly uneven with distinct extremes.
The bar for "CPR Rate \& Rhythm" is visually the tallest, reaching a count of 30. This dominance is marked with triple asterisks, indicating a statistically significant overrepresentation (p<.001).
Conversely, "CPR for Different Demographics" at the far right is among the lowest bars (count < 5). However, it is marked with a single asterisk, indicating that this scarcity is also statistically significant (p<.05), representing a notable underrepresentation compared to other topics.
Video-IVA Condition (Right Chart): The Video condition shows a different pattern. The tallest bar is "Chest Compression Techniques," reaching a count of approximately 17. This peak is marked with double asterisks, indicating statistical significance (p<.01). Unlike the AR graph, "CPR Rate \& Rhythm" is much lower here (count of 8).
}
\end{figure*}

\begin{figure*}[t]
\centering
\includegraphics[width=.9\linewidth]{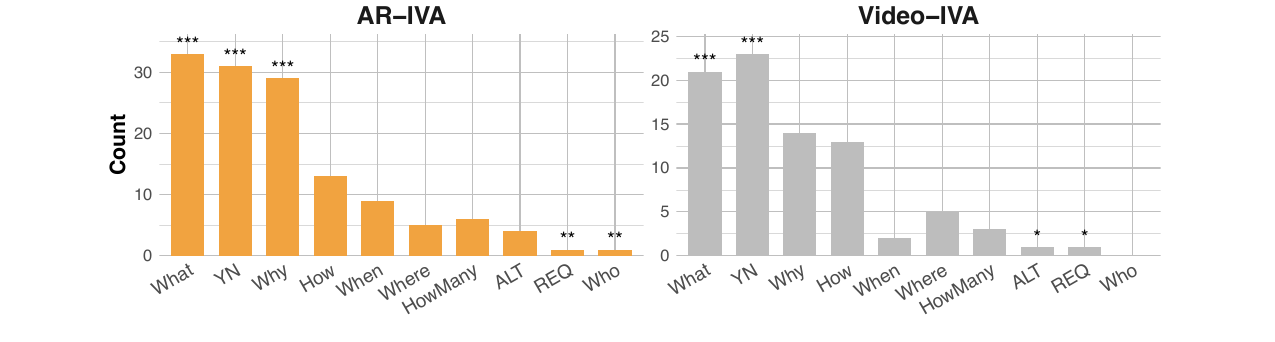}
\caption{Distributions of \emph{question forms} across AR-IVA and Video-IVA conditions. (*: $p_{adj} < 0.05$; **: $p_{adj} < 0.01$; ***: $p_{adj} < 0.001$)}
\label{fig:h2_form}
\Description{This figure presents two bar charts comparing the frequency distribution of question forms—categorized into types such as Yes/No (YN), Alternative (ALT), Requests (REQ), and WH-interrogatives—between the AR-IVA and Video-IVA conditions. The vertical axis represents the count, with asterisks indicating statistical deviations from a uniform distribution.
In the AR-IVA condition (Left Chart), the distribution is heavily skewed toward three dominant types: "What"(z=5.74), "YN" (z=5.16), and "Why" (z=4.58). These bars are visually the tallest, all exceeding a count of 28, and are marked with triple asterisks to indicate highly significant overrepresentation. Conversely, "REQ"and "Who" appear as the lowest bars (count < 2), marked with double asterisks to denote significant underrepresentation (z=−3.54).
The Video-IVA condition (Right Chart) shares the focus on basic facts, with "What" and "YN" remaining the tallest bars marked with triple asterisks . However, a key visual divergence is observed in "Why" questions; unlike the AR condition, the "Why" bar is significantly shorter (count approx. 14) and bears no asterisks, implying it was not statistically dominant. Among the rarest forms, "ALT" and "REQ" are marked with a single asterisk , indicating statistical scarcity (p=.047).
}
\end{figure*}

\textbf{What. } 
For \emph{question topic}, no significant association with condition was observed ($p = .380$), leading to a rejection of \textbf{H2-2a}. Nonetheless, as shown in Figure~\ref{fig:h2_topic}, topic distributions within each condition deviated from uniformity.

In the AR-IVA condition, the distribution of question topics was significantly non-uniform ($\chi^2(12, N = 132) = 91.17$, $p < .001$). Standardized residual analysis indicated a significant overrepresentation of ``CPR Rate and Rhythm'' ($p_{adj} < .001$, $z = 5.45$). The Video-IVA condition also showed a similarly non-uniform distribution ($\chi^2(12, N = 83) = 61.25$, $p < .001$), with questions most strongly concentrated on ``Chest Compression Techniques'' ($p_{adj} = .006$, $z = 3.61$). These patterns are consistent with AR contexts that emphasize the immediate, physical aspects of task execution, such as pacing. 
In addition, questions related to ``CPR for Different Demographics'' were significantly underrepresented, particularly in the AR-IVA condition ($p_{adj} = .039$, $z = -3.03$). 
This pattern indicates reduced attention to situational variations in immersive settings.

For \emph{question type}, no significant association with condition was observed ($p = .888$), leading to a rejection of \textbf{H2-2b}. Within-condition analyses further indicated no significant preference between \emph{procedural} and \emph{exploratory} questions in either the AR-IVA ($\chi^2(1, N = 132) = 2.45$, $p = .117$) or Video-IVA ($\chi^2(1, N = 83) = 0.98$, $p = .323$) condition. These results indicate comparable distributions of question types across conditions.

For \emph{question form}, no significant association with condition was observed ($p = .707$), leading to a rejection of \textbf{H2-2c}. Nonetheless, as shown in Figure~\ref{fig:h2_form}, form distributions within each condition deviated from uniformity.

In the AR-IVA condition, the distribution of question forms was significantly non-uniform ($\chi^2(9, N = 132) = 111.9$, $p < .001$). Standardized residual analysis indicated significant overrepresentation of ``What'' ($p_{adj} < .001$, $z = 5.74$), ``Yes/No'' ($p_{adj} < .001$, $z = 5.16$), and ``Why'' forms ($p_{adj} < .001$, $z = 4.58$), alongside significant underrepresentation of ``REQ'' ($p_{adj} = .005$, $z = -3.54$) and ``Who'' forms ($p_{adj} = .005$, $z = -3.54$). The Video-IVA condition also showed a significantly non-uniform distribution ($\chi^2(8, N = 83) = 66.1$, $p < .001$), with ``What'' ($p_{adj} < .001$, $z = 4.11$) and ``Yes/No'' ($p_{adj} < .001$, $z = 4.81$) forms overrepresented, and ``ALT'' and ``REQ'' forms underrepresented (both $p_{adj} = .047$, $z = -2.87$).

Both conditions were characterized by a predominance of ``What'' and ``Yes/No'' forms, reflecting a focus on basic clarification and confirmation. 
In contrast, the AR-IVA condition uniquely exhibited an overrepresentation of ``Why'' forms, indicating greater engagement with causal reasoning in immersive contexts.

\begin{figure*}[t]
\centering
\includegraphics[width=\linewidth]{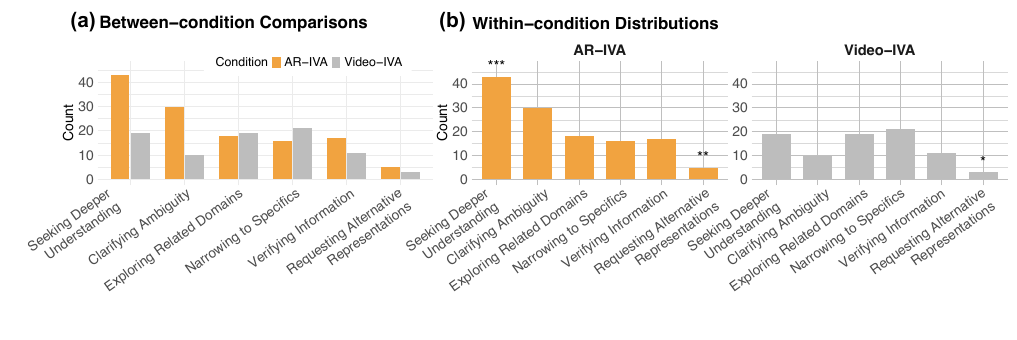}
\caption{Distributions of \emph{question intentions} across AR-IVA and Video-IVA conditions. (*: $p_{adj} < 0.05$; **: $p_{adj} < 0.01$; ***: $p_{adj} < 0.001$)}
\label{fig:h2_why}
\Description{This figure analyzes ``Question Intentions'' through two distinct lenses: between-condition comparisons in Panel (a) and within-condition distributions in Panel (b). Panel (a) visually contrasts raw counts, showing that the AR-IVA condition (Gold) trends higher in "Clarifying Ambiguity" (30 vs 10), while the Video-IVA condition (Grey) shows a higher count for "Narrowing to Specifics" (21 vs 16). Panel (b) reveals a striking difference in internal focus. The AR-IVA condition (Left) displays a highly uneven distribution, dominated by "Seeking Deeper Understanding," which is the tallest bar (count > 40), marked with triple asterisks to indicate significant overrepresentation. Conversely, Requesting Alternative Representations is significantly underrepresented in the AR-IVA condition, as indicated by double asterisks. In contrast, the Video-IVA condition (right) exhibits a much flatter distribution overall, with Requesting Alternative Representations remaining the lowest category but marked with only a single asterisk.
}
\end{figure*}

\textbf{Why. } 
For \emph{question intention}, a significant association with condition was observed ($p = .028$), supporting \textbf{H2-3}. Nonetheless, as shown in Figure~\ref{fig:h2_why}(a), post-hoc pairwise comparisons with FDR correction revealed no individual intention categories that differed significantly between conditions ($p_{adj} > .05$ for all).

Exploratory comparisons using unadjusted p-values suggested potential sources of this overall association. Learners in the AR-IVA condition tended to exhibit more ``Clarifying Ambiguity'' intentions (AR: 30 vs. Video: 10; $p = .048$), whereas learners in the Video-IVA condition more frequently showed ``Narrowing to Specifics'' intentions (AR: 16 vs. Video: 21; $p = .025$). Although the ``Seeking Deeper Understanding'' intention appeared more frequently in the AR-IVA condition (32.6\%) than in the Video-IVA condition (22.9\%), this difference was not statistically significant ($p = .12$). This suggests that the higher raw count may be attributable to the greater overall question volume rather than a proportional shift in intention.

In the AR-IVA condition, the distribution of question intentions was significantly non-uniform ($\chi^2(5, N = 129) = 40.44$, $p < .001$). As shown in Figure~\ref{fig:h2_why}(b), Standardized residual analysis indicated a significant overrepresentation of ``Seeking Deeper Understanding'' ($p_{adj} < .001$, $z = 5.08$) and an underrepresentation of ``Requesting Alternative Representations'' ($p_{adj} = .002$, $z = -3.90$). The Video-IVA condition showed a similarly non-uniform distribution ($\chi^2(5, N = 83) = 17.70$, $p = .003$), with ``Requesting Alternative Representations'' significantly underrepresented ($p_{adj} = .021$, $z = -3.19$). These patterns indicate that learner questions in the AR-IVA condition were more strongly oriented toward deeper conceptual understanding of the IVA’s feedback.

\begin{figure}[t]
\centering
  \begin{subfigure}[t]{0.4\linewidth}
    \centering
    \includegraphics[width=\linewidth]{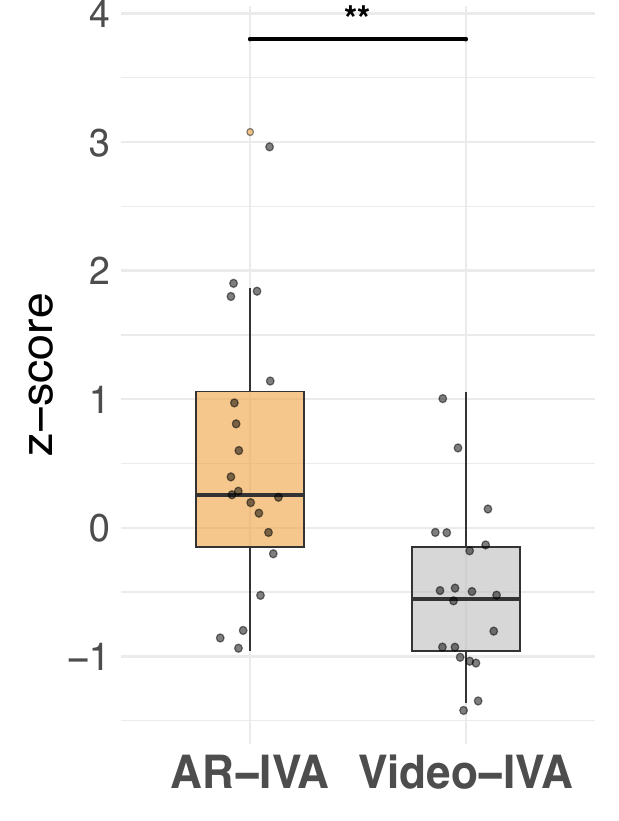}
    \caption{Query count}
    \label{fig:query_count}
  \end{subfigure}
  \begin{subfigure}[t]{0.4\linewidth}
    \centering
    \includegraphics[width=\linewidth]{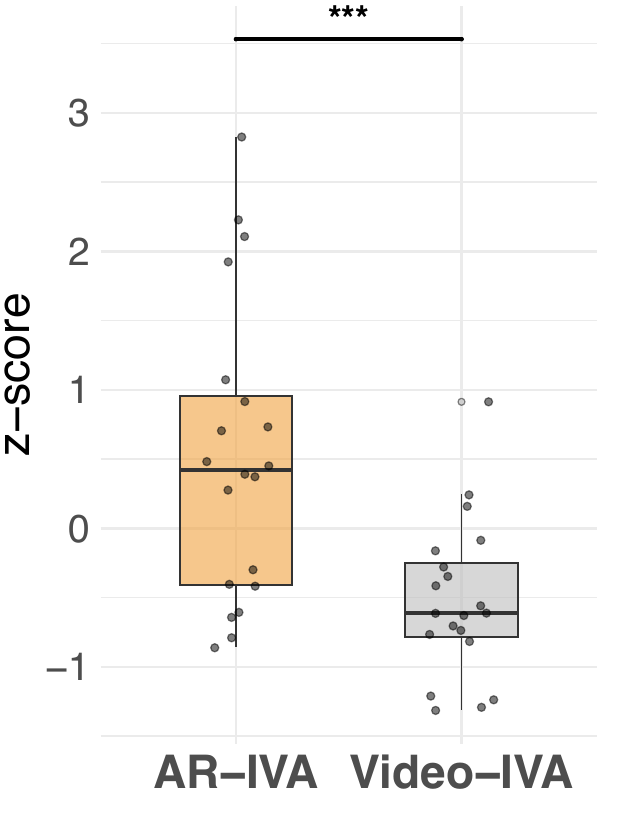}
    \caption{Query length}
    \label{fig:query_length}
  \end{subfigure}
\caption{Effects of presentation modality on the components of \emph{query volume} ( **: $p < 0.01$; ***: $p < 0.001$)}
\label{fig:h2_4}
\Description{This figure illustrates the effect of presentation modality on query volume using side-by-side box plots based on standardized z-scores, where the 0 line represents the sample average. Visually, the charts reveal a distinct contrast where the AR-IVA condition (Gold) is consistently elevated above the mean, while the Video-IVA condition (Grey) falls below it.
The Left Chart for Query Count shows that AR learners asked significantly more questions compared to the video group, a difference marked by double asterisk (p=.001).
The Right Chart for Query Length displays an even stronger pattern, where the AR condition is positioned entirely above the baseline. This highly significant difference is highlighted by triple asterisks, confirming that the immersive environment encouraged learners to ask not only more frequent but also lengthier questions.
}
\end{figure}


\textbf{How. }  
For \emph{query volume}, a significant association with condition was observed. 
As shown in Figure~\ref{fig:h2_4}, for \emph{query count}, the AR-IVA condition showed higher standardized scores ($M = 0.49$) than the Video-IVA condition ($M = -0.49$), $t(30.86) = 3.57$, $p = .001$. 
Similarly, for \emph{query length}, the AR-IVA condition also exhibited significantly higher values ($M = 0.52$) than the Video-IVA condition ($M = -0.52$), $t(28.66) = 3.85$, $p < .001$. 
Thus, \textbf{H2-4a} was supported.

By contrast, for \emph{query complexity}, no significant differences were observed  ($p = .844$), leading to a rejection of \textbf{H2-4b}. These results indicate that the AR-IVA condition was associated with higher query frequency and length, whereas no significant differences were found for query complexity.

\subsubsection{Mediation Effect of Presence}

\begin{figure*}[t]
\centering
 \includegraphics[width=\linewidth]{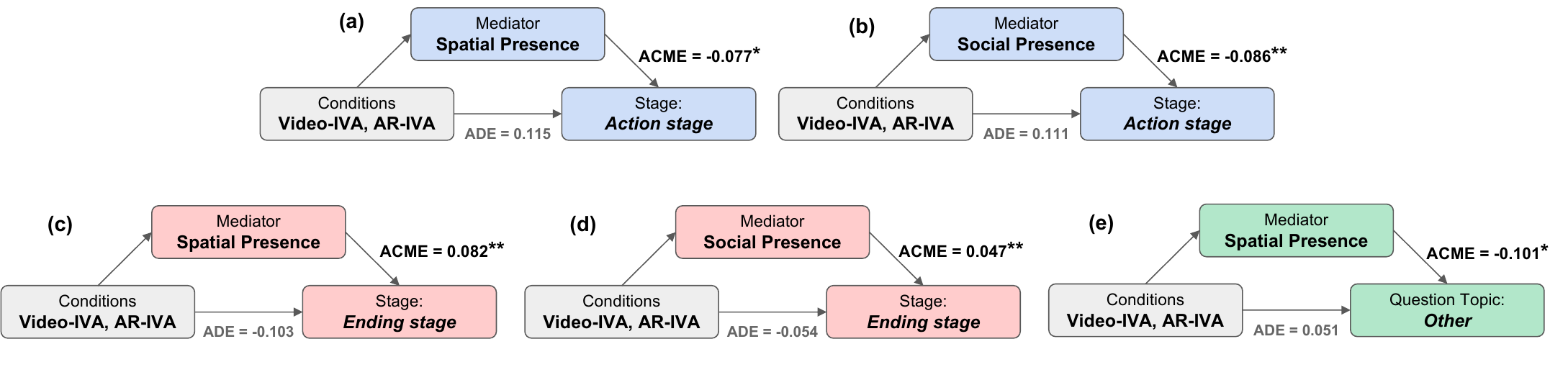}
\caption{Mediation effects of presence on \emph{question timing} (a--d) and \emph{question topic} (e). 
(*: $p_{adj} < 0.05$; **: $p_{adj} < 0.01$).}
\label{fig:H3-1_H3-2}
\Description{This figure illustrates five mediation path diagrams analyzing how presence influences questioning behavior across different stages. Each diagram connects "Conditions" (Video-IVA vs. AR-IVA) to a "Mediator" (Spatial/Social Presence), and finally to an "Outcome" (Stage or Topic).
Figure (a) Spatial Presence → Action Stage: The arrow connecting the mediator (Spatial Presence) to the "Action stage" carries a negative coefficient (ACME = -0.077), marked with a single asterisk. This visually indicates a negative relationship during the action phase.
Figure (b) Social Presence → Action Stage: Similarly, the arrow from "Social Presence" to the "Action stage" displays a negative coefficient (ACME = -0.086), marked with double asterisks.
Figure (c) Spatial Presence → Ending Stage: In contrast, the arrow connecting "Spatial Presence" to the "Ending stage" carries a positive coefficient (ACME = 0.082), marked with double asterisks, indicating a positive effect.
Figure (d) Social Presence → Ending Stage: The arrow from "Social Presence" to the "Ending stage" also displays a positive coefficient (ACME = 0.047), marked with double asterisks.
Figure (e) Spatial Presence → Question Topic 'Other': The arrow connecting "Spatial Presence" to the outcome "Question Topic: Other" carries a negative coefficient (ACME = -0.101), marked with a single asterisk.
}
\end{figure*}
 
\textbf{When.} 
For \emph{question timing}, the analysis revealed no significant mediation effects of spatial presence ($ACME = 0.006$, $p_{adj} = .860$) and social presence ($ACME = 0.039$, $p_{adj} = .070$) during the \emph{Explanation} stage. 
During the \emph{Action} stage, however, both spatial presence ($ACME = -0.077$, $p_{adj} = .042$) and social presence ($ACME = -0.086$, $p_{adj} = .006$) exhibited significant negative indirect effects (Figure~\ref{fig:H3-1_H3-2}(a)--(b)). 
In contrast, during the \emph{Ending} stage, both spatial presence ($ACME = 0.082$, $p_{adj} = .006$) and social presence ($ACME = 0.047$, $p_{adj} = .006$) showed significant positive mediation effects (Figure~\ref{fig:H3-1_H3-2}(c)--(d)). 
These results indicate that the presence differentially mediated questioning behavior across instructional stages. 
Specifically, presence was associated with suppressed questioning during active demonstration and increased reflective inquiry at the end of the session.
Accordingly, \textbf{H3-1} was partially supported.

\textbf{What. } 
For \emph{question topic}, spatial presence exhibited a significant negative mediation effect on the ``Other'' category ($ACME = -0.101$, $p_{adj} = .022$; Figure~\ref{fig:H3-1_H3-2}(e)). Accordingly, \textbf{H3-2a} was partially supported. 
In contrast, no significant mediation effects were observed for \emph{question type} or \emph{question form} ($p_{adj} > .05$). Accordingly, \textbf{H3-2b} and \textbf{H3-2c} were rejected.
This finding indicates a selective mediation effect of spatial presence on specific topic categories.

\textbf{Why. }  
The analysis revealed no significant mediation effects of either spatial or social presence on \emph{question intention} ($p_{adj} > .05$), leading to a rejection of \textbf{H3-3}. However, as reported in the H2-3 results, presentation modality itself was significantly associated with differences in question intention. These findings indicate that although presentation modality influenced learners’ question intention, this effect was not mediated by presence.

\textbf{How. }  
The analysis revealed no significant mediation effects of spatial or social presence on \emph{query volume} (i.e., \emph{query count} and \emph{query length}) or on \emph{query complexity} ($p_{adj} > .05$), leading to a rejection of \textbf{H3-4}. However, as reported in the H2-4a results, presentation modality itself had a significant effect on query volume. These findings indicate that although presentation modality influenced learners’ query volume, this effect was not mediated by presence.

\subsubsection{Moderating Effect of Learning Motivation and Strategies}

\begin{figure*}[t]
\centering
\includegraphics[width=\linewidth]{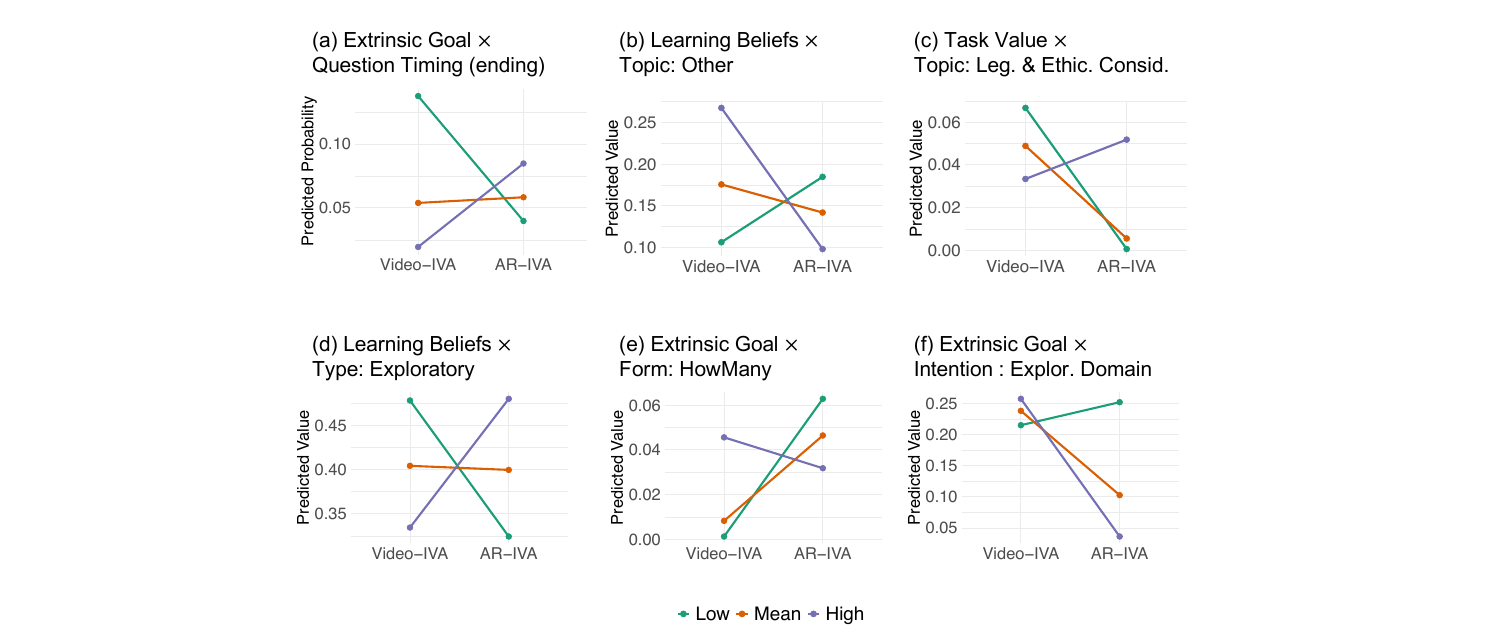}
\caption{Moderation effects of learning motivation on \emph{question timing} (a),
\emph{question topic} (b--c), \emph{question type} (d), \emph{question form} (e), and \emph{question intention} (f). Abbreviations: Leg. \& Ethic. Consid. = Legal and Ethical Considerations; Explor. Domain = Exploring Related Domain.} 
\label{fig:h4-1}
\Description{
This figure presents a series of interaction plots illustrating the moderating effects of motivational traits (Low, Mean, High) across the Video-IVA and AR-IVA conditions. The plots map the "Predicted Value" on the y-axis against the two conditions on the x-axis.
Figure (a) L Extrinsic Goal × Question Timing (ending): The plot reveals a distinct divergence. The line for the High group (Purple) shows a steep upward slope from Video to AR, indicating an increase in questions in the ending stage. In contrast, the Low group (Green) shows a slight downward trend.
Figure (b) Learning Beliefs × Topic 'Other': The plot reveals a distinct divergence. The line for the High group (Purple) shows a steep downward slope from Video to AR, indicating a sharp decrease in 'Other' questions. In contrast, the Low group (Green) shows a slight upward trend.
Figure (c) Task Value × Topic 'Legal \& Ethic': A crossover interaction is visible. The High group (Purple) line slopes upward, showing an increase in legal/ethical questions in the AR condition, whereas the Low group (Green) line slopes sharply downward.
Figure (d) Learning Beliefs × Type 'Exploratory': The High group (Purple) line rises sharply, crossing over the Mean and Low lines, which remain relatively flat or slope downward. This visually highlights that those with high learning beliefs increased their exploratory questions in the AR setting.
Figure (e) Extrinsic Goal × Form 'HowMany': The lines exhibit a fanning pattern. The Low group (Green)shows a distinct upward spike in the AR condition, while the High group (Purple) remains flat and low. This suggests that low extrinsic motivation led to more 'HowMany' questions.
Figure (f) Extrinsic Goal × Intent 'Explor. Domain': This plot shows a clear "X-shaped" crossover. The High group (Purple) slopes steeply downward, while the Low group (Green) slopes upward. This visualizes a complete reversal in behavior between the two groups depending on the condition.
}
\end{figure*}

\paragraph{Learning Motivation}
 
\textbf{When. }
For \emph{question timing}, the analysis revealed a significant moderation effect of extrinsic goal orientation. Specifically, as shown in Figure~\ref{fig:h4-1}(a), learners with higher extrinsic motivation showed a significant increase in questioning during the \emph{Ending} stage within the AR-IVA condition ($b = 1.57, SE = 0.66, p = .017$). 
This indicates that learners driven by external goals tended to ask their questions during the ending stage, using it as a reflective phase following the core instruction. Consequently, \textbf{H4-1a} was partially supported.


\textbf{What. }
For \emph{question topic}, as shown in Figure~\ref{fig:h4-1}(b), learners with stronger control of learning beliefs, who believe their efforts directly determine learning outcomes, showed a steep decrease in peripheral ``Other'' questions within the AR-IVA condition ($b = -2.94$, $SE = 1.27$, $p = .020$). This indicates a tendency to self-regulate and filter out irrelevant questions to focus on core content. Conversely, as shown in Figure~\ref{fig:h4-1}(c), those with higher task value, who perceive the learning material as interesting and useful, were more likely to raise questions regarding ``Legal and Ethical Considerations'' in the AR-IVA condition ($b = 4.20$, $SE = 1.89$, $p = .026$). This suggests that attributing high value to the task motivated them to explore its deeper contextual implications beyond immediate procedures. 
These results collectively suggest that learners’ motivational traits were associated with the topical focus of questioning.

For \emph{question type}, as shown in Figure~\ref{fig:h4-1}(d), learners with stronger control of learning beliefs were more likely to ask exploratory questions in the AR-IVA condition ($b = 0.93$, $SE = 0.39$, $p = .017$). 
These results indicate that learners’ motivational traits were associated with differences in question type.

For \emph{question form},  as shown in Figure~\ref{fig:h4-1}(e), learners with higher extrinsic goal orientation showed a decrease in the use of ``HowMany'' forms within the AR-IVA condition ($b = -3.33$, $SE = 1.57$, $p = .034$). This indicates reduced reliance on simple quantitative checks and suggests that question form is associated with learners’ motivational profiles. Taken together, \textbf{H4-1b} was partially supported across question topic, type, and form.

\textbf{Why. }  
For \emph{question intention}, as shown in Figure~\ref{fig:h4-1}(f), learners with higher extrinsic goal orientation showed a decrease in the “Exploring Related Domain” intention within the AR-IVA condition ($b = -1.42$, $SE = 0.53$, $p = .007$). This indicates that learners’ motivational traits were associated with differences in question intention. Thus, \textbf{H4-1c} was partially supported.

\textbf{How. }  
The analysis revealed no significant mediation effects of motivational traits on \emph{query volume} (i.e., \emph{query count} and \emph{query length}) or on \emph{query complexity} ($p > .05$), leading to a rejection of \textbf{H4-1d}. These results suggest that learners’ motivational traits did not alter the amount or complexity of questions posed across conditions.

\paragraph{Learning Strategies} 
\begin{figure*}[t]
\centering
\includegraphics[width=\linewidth]{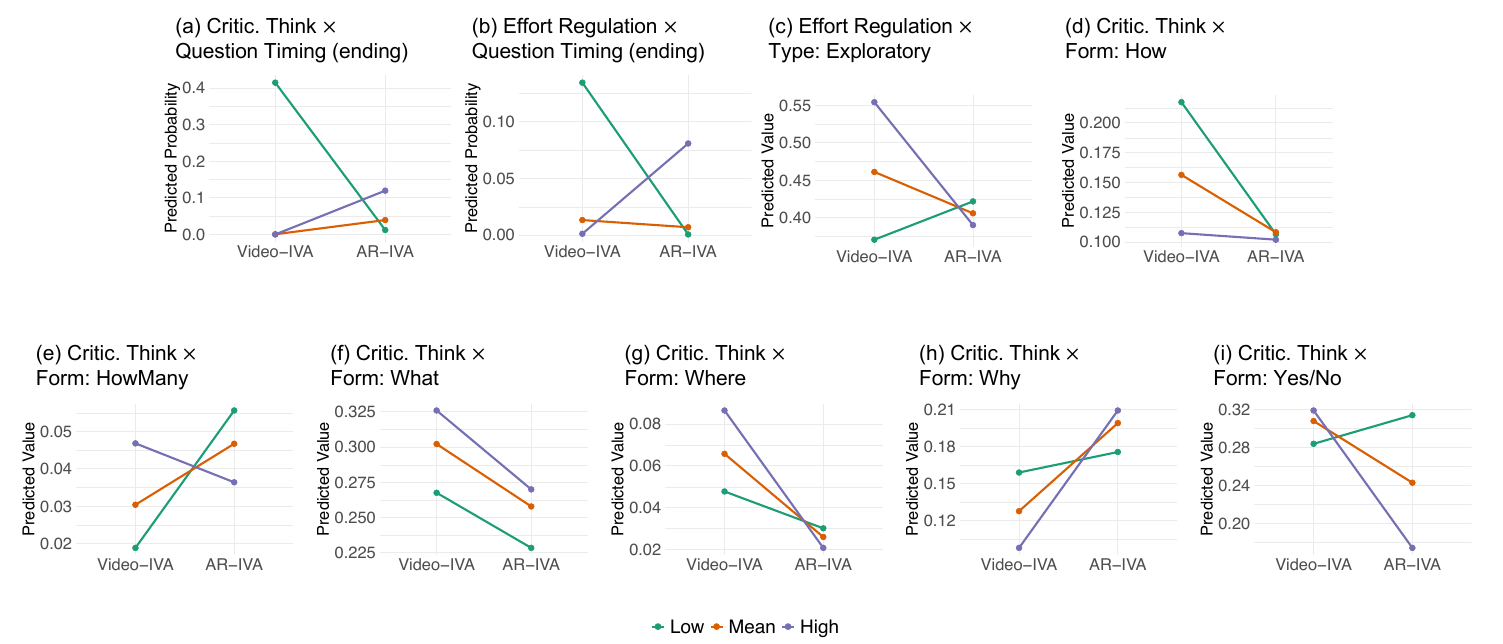}
\caption{Moderation effects of learning strategies on \emph{question timing} (a--b), \emph{question type} (c), and \emph{question form} (d--i). Abbreviations: Critic. Think. = Critical Thinking.}
\label{fig:h4_2_1}
\Description{
This figure presents three interaction plots illustrating how learning strategies moderate learner questioning behavior across conditions. The plots map the "Predicted Value" on the y-axis against the Video-IVA and AR-IVA conditions on the x-axis, comparing learners with High (Purple) versus Low (Green) levels of each strategy.
Figure (a), labeled ``Critic. Think. x Question Timing (ending),'' focuses on Critical Thinking. It displays a clear crossover pattern where the line for the High group (Purple) slopes upward from Video to AR, indicating increased questioning at the end, whereas the Low group (Green) slopes downward.
Figure (b), labeled ``Effort Regulation x Question Timing (ending),'' shows a similar trend for Effort Regulation. Here, the High group (Purple) again slopes upward, suggesting that learners with strong regulation also delayed questions to the end in the AR setting, while the Low group (Green) shows a sharp downward drop.
Figure (c), labeled ``Effort Regulation x type: Exploratory,'' reveals a reversed crossover. The High Effort Regulation group (Purple) follows a steep downward slope, suggesting a decrease in exploratory questions within the AR environment. Conversely, the Low group (Green) shows an upward trend, indicating they asked more exploratory questions in the immersive setting.
This figure presents a series of interaction plots illustrating how "Critical Thinking" (Low=Green, High=Purple) moderates the frequency of different "Question Forms" across Video-IVA and AR-IVA conditions.
Figure (d) "Critic. Think. x form: How": Shows a convergence pattern. The Low group (Green) starts high in the Video condition but follows a steep downward slope. The High group (Purple) also slopes downward but more gently, resulting in both groups meeting at a low point in the AR condition.
Figure (e) "Critic. Think. x form: HowMany": Displays a clear crossover. The Low group (Green) slopes steeply upward, indicating an increase in 'HowMany' questions in AR. Conversely, the High group (Purple)slopes downward.
Figure (f) "Critic. Think. x form: What": Shows parallel downward trends. Both groups decrease in 'What' questions in the AR condition, but the High group (Purple) consistently remains higher than the Low group (Green) throughout.
Figure (g) "Critic. Think. x form: Where": Similar to (a), this shows a convergence. The High group (Purple)starts as the highest point but drops sharply to meet the Low group (Green), which also declines. This indicates that variance between groups disappears in the AR setting.
Figure (h) "Critic. Think. x form: Why": Shows a dramatic divergence. The High group (Purple) starts as the lowest point in the video but follows a steep upward slope to become the highest in AR. The Low group (Green)remains relatively flat. This visually confirms that highly critical thinkers specifically increased 'Why' questions in the immersive environment.
Figure (i) "Critic. Think. x form: Yes/No": Displays a steep crossover. The High group (Purple) starts high but plunges steeply downward in the AR condition. In contrast, the Low group (Green) slopes upward, indicating they relied more on simple Yes/No confirmations in the AR setting.

}
\end{figure*}

\textbf{When. } 
For \emph{question timing}, as shown in Figure~\ref{fig:h4_2_1}(a--b), learners with higher critical thinking ($b = 6.52$, $SE = 2.37$, $p = .006$) and higher effort regulation ($b = 5.38$, $SE = 1.36$, $p < .001$) showed a significant increase in questioning during the \emph{Ending} stage in the AR-IVA condition. These results indicate that learners with stronger cognitive and regulatory skills strategically concentrated their questioning during the ending stage. Consequently, \textbf{H4-2a} was partially supported.

\textbf{What. }  
For \emph{question topic}, no significant moderation effects of learning strategies were observed.
For \emph{question type}, as shown in Figure~\ref{fig:h4_2_1}(c), learners with higher effort regulation showed a decrease in exploratory questions in the AR-IVA condition ($b = -0.81$, $SE = 0.37$, $p = .028$). These results indicate that learners’ self-regulatory capacity was associated with differences in question type.
For \emph{question form}, as shown in Figure~\ref{fig:h4_2_1}(d--i), critical thinking significantly moderated multiple question form subtypes in the AR-IVA condition, including \emph{How}, \emph{HowMany}, \emph{What}, \emph{Where}, \emph{Why}, and \emph{Yes/No} (all $p < .001$). These results indicate that critical thinking was broadly associated with variation in question form. 
Collectively, these findings highlight the significant impact of learning strategies, particularly critical thinking and effort regulation, on the type and form of questions. Thus, \textbf{H4-2b} was partially supported. 

\begin{figure*}[t]
\centering
\includegraphics[width=\linewidth]{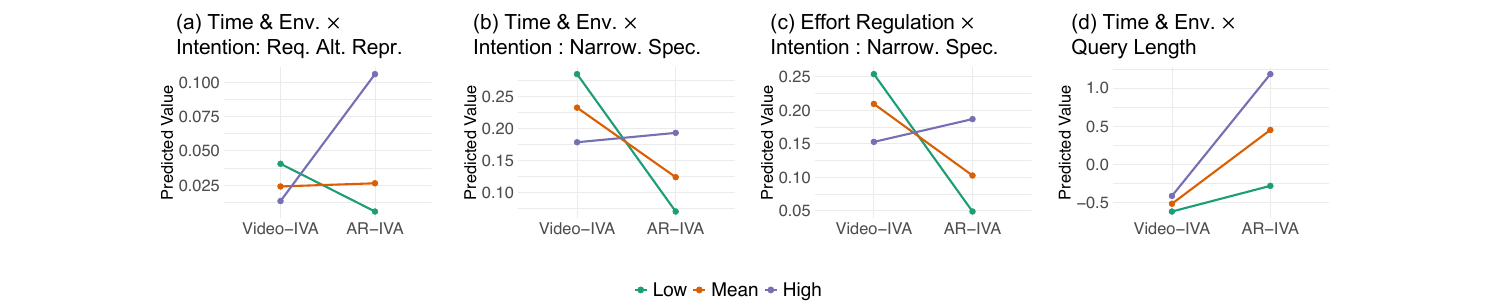}
\caption{Moderation effects of learning strategies on \emph{question intention} (a--c) and \emph{query length} (d). Abbreviations: Time \& Env. = Time and Study Environment Management; Req. Alt. Repr. = Requesting Alternative Representations; Narrow. Spec. = Narrowing to Specifics.}
\label{fig:h4_3}
\Description{This figure illustrates the moderation effects of learning strategies (Time \& Study Environment, Effort Regulation) on "Question Intention" and "Query Length".
Figure (a) "Time \& Env. x intent: Req. Alt. Repr.": Shows a crossover interaction. The High group (Purple)slopes steeply upward, indicating they sought different representations more often in AR. The Low group (Green )slopes downward.
Figure (b) "Time \& Env. x intent: Narrow. Spec. ": Displays a crossover. The Low group (Green) starts at the highest point in Video but drops steeply downward in AR. The High group (Purple) shows a slight upward trend, becoming higher than the Low group in the AR condition.
Figure (c) "Effort Regulation x intent: Narrow. Spec. ": Also shows a crossover. Similar to (b), the Low group (Green) slopes steeply downward, while the High group (Purple) slopes upward, indicating that learners with better effort regulation increased their narrowing-down questions in the immersive setting.
Figure (d) "Time \& Env. x Query Length": Shows a significant divergence. The High group (Purple) exhibits a dramatic, steep upward slope, indicating a massive increase in question length in the AR condition. The Low group (Green) shows only a very slight increase, remaining much lower.
}
\end{figure*}

\textbf{Why. }  
For \emph{question intention}, as shown in Figure~\ref{fig:h4_3}(a--c), learners with stronger time and study environment management showed an increase in ``Requesting Alternative Representations'' ($b = 2.83$, $SE = 1.11$, $p = .011$) and ``Narrowing to Specifics'' ($b = 1.15$, $SE = 0.55$, $p = .038$) intentions in the AR-IVA condition. Similarly, learners with higher effort regulation exhibited an increase in the ``Narrowing to Specifics'' intention ($b = 1.38$, $SE = 0.68$, $p = .042$) in the AR-IVA condition. These results indicate that learners with stronger self-regulation tended to engage in more refined and targeted inquiry intentions in immersive settings. Consequently, \textbf{H4-2c} was partially supported. 

\textbf{How. }  
For \emph{query volume}, as shown in Figure~\ref{fig:h4_3}(d), time and study environment management significantly moderated \emph{query length} ($b = 0.63$, $SE = 0.25$, $t = 2.50$, $p = .017$), whereas no significant moderation effects were observed for \emph{query count} and \emph{query complexity}. These results indicate that learners with stronger time and study environment management skills tended to formulate longer questions. Consequently, \textbf{H4-2d} was partially supported.

\begin{figure}[t]
\centering
\includegraphics[width=.6\linewidth]{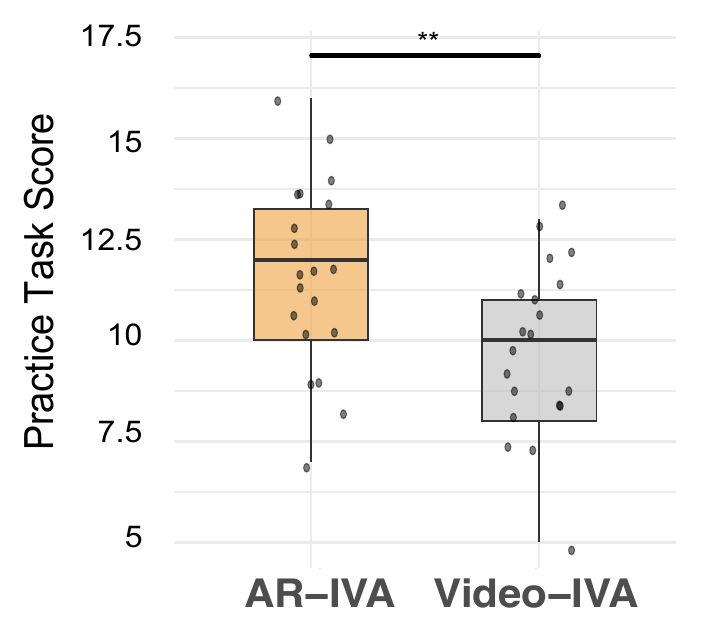}
\caption{Effects of presentation modality on \emph{CPR practical performance evaluation} ($**:p<0.01$)}
\label{fig:h5}
\Description{This figure illustrates the effect of presentation modality on learning outcomes, specifically ``Practice Task Performance.'' The box plots reveal a distinct difference in vertical positioning between the two groups. The AR-IVA box (Gold) is elevated significantly higher on the ``Practice Task Score'' axis compared to the Video-IVA box (Grey), which is situated lower. This visual gap indicates that the AR group achieved superior performance scores. A bracket connecting the two conditions is marked with double asterisks, visually confirming that the AR-IVA condition significantly outperformed the Video-IVA condition in the practical application of skills.
}
\end{figure}

\subsubsection{Effect of Modality on Learning Outcome} 
No significant difference was observed in knowledge quiz scores between the AR-IVA condition ($M = 7.75$, $SD = 0.55$) and the Video-IVA condition ($M = 7.85$, $SD = 0.37$), $t(38) = -0.68$, $p = .503$. 
By contrast, as shown in Figure~\ref{fig:h5}, practical performance scores were significantly higher in the AR-IVA condition ($M = 11.65$, $SD = 2.37$) than in the Video-IVA condition ($M = 9.70$, $SD = 2.13$), $t(38) = 2.74$, $p = .009$. 
These results indicate that while AR-IVA did not enhance declarative knowledge, the AR-IVA condition yielded higher practical performance scores, reflecting differences in procedural learning outcomes across modalities. Accordingly, \textbf{H5} was partially supported.

\subsection{Participant Comments} 

We conducted semi-structured interviews to gather qualitative insights into how presentation modality shaped participants’ questioning behavior. 
After classifying their question intentions, participants elaborated on the reasons behind their questions in the interviews. 
Interview transcripts were analyzed using thematic analysis~\cite{joffe2011thematic, riger2016thematic}. 
Two researchers independently examined the transcripts to identify emerging patterns related to spatial and social presence, motivational factors, and questioning strategies. 
Discrepancies were resolved through discussion until consensus was reached, ensuring the trustworthiness and reliability of the qualitative findings. 
This analytic process informed the themes reported below.

\textbf{Dual Role of Presence: Shared Comfort, Divergent Engagement Effects. }
Interview results indicated that learners in both conditions perceived the avatar-based IVA as less socially pressuring than a human instructor. This reduced social pressure made it easier for learners to ask questions. One participant noted, ``There are not many opportunities for one-on-one lessons… when other students are present, but here I could ask anytime without feeling pressure'' (P3, AR). Another explained, ``When asking a human instructor, I feel self-conscious about asking the same thing multiple times, but here I could immediately clarify and move on'' (P15, AR). Similarly, another participant added, ``With a real instructor I was too shy to ask, but with the agent it was easier'' (P22, Video). These comments suggest that the avatar-based IVA provided a sense of psychological comfort.

However, the way presence shaped learners’ engagement differed by condition. 
In the AR-IVA condition, the avatar created a form of positive social pressure that helped sustain learners’ attention during instruction. 
As one participant remarked, ``With AR, there was an avatar in front of me, so if I lost focus, I felt guilty, which helped me stay attentive'' (P4, AR). 
Another stated, ``It felt like the avatar was actually right in front of me, which made me feel that I had to pay attention to the class'' (P7, AR). 

In contrast, in the Video-IVA condition, although the IVA provided content identical to that of the AR-IVA condition, the weaker sense of social presence limited the agent’s perceived authority and credibility. 
As one participant noted, ``Q\&A was useful, but the answers lacked credibility and left questions unresolved'' (P25, Video). 
Another added, ``It was distracting that the agent kept giving similar responses'' (P23, Video). 
These findings suggest that while a strong presence in AR helped sustain learners’ focus, a limited presence in the video condition undermined trust in the agent and reduced engagement.

\textbf{Questioning Patterns: Conceptual Deepening vs. Procedural Refinement. } 
Across both the AR-IVA and Video-IVA conditions, learners asked questions aimed at understanding and confirming CPR procedures.
These included basic procedural checks as well as explanatory or context-related inquiries, indicating broadly similar question types across conditions.

However, the underlying questioning strategies diverged across conditions.
In the AR-IVA condition, learners tended to engage in conceptually deeper questioning, extending beyond procedural clarification to explore underlying reasoning and contextual understanding.
For example, one participant explained, ``I asked, `Why do compressions have to be vertical?' to understand the rationale behind the procedure'' (P1, AR).
Similarly, another participant stated, ``While performing CPR, I was curious about what to do if a bone were fractured and whether I would be held responsible for it'' (P6, AR), while another noted, ``The procedure was clear, so I focused more on asking why'' (P4, AR).

By contrast, in the Video-IVA condition, questioning strategies were primarily oriented toward narrowing inquiries to specific procedural details and achieving procedural precision.
For example, one participant stated, ``I asked to check the detailed procedure'' (P15, Video), while another noted, ``I asked about exceptional situations that could occur'' (P27, Video).
Another participant explained, ``I asked about procedures because I was not fully certain about the answer'' (P25, Video).
Together, these comments suggest that presentation modality shaped learners’ approaches to managing uncertainty during learning. The embodied and immersive nature of AR encouraged learners to treat procedural knowledge as a starting point for deeper conceptual reasoning, whereas the screen-based Video-IVA led learners to resolve uncertainty by refining and confirming specific procedural steps.

\section{Discussion}
This section interprets the key findings of the study and discusses their broader implications. We first analyze how the results address our research hypotheses regarding the effects of presentation modality, presence, and individual differences on learners' questioning behavior. Based on this analysis, we propose a set of design implications for developing more effective and personalized learning systems. The section concludes with an acknowledgment of the study's limitations and suggests directions for future research.

\subsection{Analysis on the Study Results}

\textbf{Presence as a Foundational Affordance (H1). }
The AR-IVA condition elicited significantly higher levels of spatial and social presence than the Video-IVA condition. This result confirms that presentation modality directly influences learners’ sense of presence, providing empirical support for our foundational premise (H1).  This finding positions presence as a key psychological affordance that shapes learners’ cognitive focus and subsequent questioning behavior.

\textbf{Modality Shaping Intention and Intensity (H2). } 
Across modalities, learners exhibited broadly comparable distributions in question timing and content, including topic, type, and form (H2-1, H2-2). 
However, presentation modality was significantly associated with differences in questioning intention and intensity. 
Regarding question intention (H2-3), learners in the AR-IVA condition demonstrated a proactive inquiry pattern, characterized by questions aimed at seeking deeper understanding or clarifying ambiguities in the agent’s responses.
In contrast, those in the Video-IVA condition focused on narrowing their inquiries to resolve specific, factual uncertainties.
In addition, learners in the AR-IVA condition produced a higher number of questions and longer questions compared to the Video-IVA condition (H2-4a), indicating a greater intensity of engagement.
These findings reveal that presentation modality shaped not the surface characteristics of questions, but rather the underlying intentions and the intensity with which learners engaged in questioning.

\textbf{Presence as a Selective Filter (H3). }
Despite its direct effect, presence did not function as a universal mediator linking modality to all aspects of questioning. Instead, it operated as a selective cognitive filter or spotlight. High presence suppressed questioning during the action stage, allowing learners to remain focused on the task demonstration. In contrast, during the ending stage, presence facilitated reflective questioning as learners integrated the overall instruction (H3-1). At the same time, spatial presence directed attention toward core CPR procedures while filtering out peripheral topics, functioning as a focusing mechanism (H3-2a). However, presence did not significantly alter the structural or qualitative aspects of questioning, including question type (H3-2b), form (H3-2c), intention (H3-3), or the overall intensity and depth of inquiry (H3-4). These findings indicate that presence alone cannot explain all presentation modality-based differences.

\textbf{Motivation as a Compass for Questioning (H4-1). }
Motivational traits significantly shaped question timing, content, and intention.
In the AR-IVA condition, higher extrinsic goal orientation significantly increased questioning during the ending stage (H4-1a).
It was also associated with fewer simple fact-checking questions (e.g., ``how many'') (H4-1b) and less broad domain exploration (H4-1c). These patterns indicate that extrinsic goal orientation primarily modulated when questions were asked and what scope they addressed.
In contrast, learners with stronger control of learning beliefs filtered out peripheral topics and utilized more exploratory question types. Meanwhile, those who placed higher task value expanded their question topics to broader contextual implications, such as legal and ethical considerations.
However, motivational traits did not influence question intensity or depth (H4-1d). 
Taken together, these results suggest that motivation shaped the direction and focus of inquiry, rather than the amount or elaboration of questioning.

\textbf{Strategies as a Toolkit for Execution (H4-2). }
Whereas motivation oriented learners’ inquiries, learning strategies governed their execution.
Critical thinking emerged as the most powerful moderator, influencing not only question timing (H4-2a) but also multiple grammatical forms (H4-2b), suggesting its role as a communicative strategy. Other strategies also played distinct roles. 
Effort regulation reduced exploratory questioning in the AR-IVA condition (H4-2b), while time and study environment management specifically predicted increased question length (H4-2d). Furthermore, these strategies influenced questioning intentions, amplifying modality-based differences in whether learners reframed information or refined the scope of their inquiries (H4-2c). These results indicate that learning strategies determined how learners structured and articulated their inquiries.

\textbf{Procedural Learning Advantages of AR-IVA (H5). }
The influence of presentation modality extended to learning outcomes. While no significant differences were observed between the two conditions in the acquisition of declarative knowledge (quiz scores), learners in the AR-IVA condition significantly outperformed those in the Video-IVA condition on the practice task. This suggests that the AR environment was particularly effective for supporting procedural application. This performance advantage may stem from the heightened presence during the action stage, which facilitated sustained attention to the task demonstration. Taken together, these findings suggest that AR-IVA demonstrates distinct advantages in skill-based learning contexts where the emphasis is not merely on knowing what, but on knowing how.

This study demonstrates that the presentation modality of educational IVAs does not provide a uniform experience across learners. Rather, questioning behavior is shaped by the interaction between modality and learners’ motivational orientations and strategic characteristics. These findings suggest that the effectiveness of instructional technologies should be understood not only in terms of their affordances, but also in relation to individual learner characteristics. Consequently, designing adaptive, learner-centered systems that explicitly account for such individual differences is essential.

\subsection{Design Implications}
Building on these findings, we propose three design principles for avatar-based IVA systems to optimize the interplay between presentation modality and learner characteristics.

First, avatar-based IVAs should be designed as \textbf{Modality-Adaptive Agents} that tailor interaction strategies to the affordances of the presentation modality. Our results indicate that AR promotes conceptual deepening while Video encourages procedural refinement, reflecting modality-specific information-seeking strategies. Accordingly, IVA response logic should be differentially optimized across modalities:

\begin{itemize}
\item \textbf{Alignment Strategy (Reinforcing Strengths):} The system should reinforce the learning strategies naturally induced by each medium. In AR environments, priority should be given to deep explanatory feedback focused on causal reasoning (``Why'') and conceptual connections. In Video environments, the core function should emphasize procedural verification and condition-checking feedback to ensure step-by-step accuracy.

\item \textbf{Bridging Strategy (Balancing Inquiry Focus):}  The system should also complement inquiry dimensions that are less naturally elicited by each modality. 
In AR, this involves supporting procedural accuracy alongside learners’ tendency toward conceptual deepening. 
In Video, the system should prompt reflective ``Why''-based questions to counterbalance learners’ focus on resolving factual uncertainties.
\end{itemize}

Second, IVA systems should provide \textbf{Adaptive Personalization} by dynamically adjusting interaction logic based on learners’ motivational and strategic profiles. Our findings demonstrate that these characteristics moderate questioning behavior across modalities, indicating that personalization must account for the interaction between learner traits and the specific affordances of the presentation modality:

\begin{itemize}
\item \textbf{Flexible Content Structuring:} The system should organize instructional content into flexible layers whose emphasis can be adaptively adjusted based on learners’ motivational orientations and strategic tendencies across modalities.  
Differences in motivation and learning strategies influence whether learners focus on the conceptual core or broaden and reframe their inquiries. This suggests that the prominence and sequencing of content layers should remain dynamically adjustable rather than fixed.

\item \textbf{Adaptive Scaffolding of Inquiry Execution:} The systems hould adapt scaffolding to both learner characteristics and interaction context.  
This may involve lightweight prompts or guided inquiry cues when exploratory or reflective questioning is attenuated, supporting productive inquiry without enforcing uniform conversational rules.

\item \textbf{Robust NLU for High-Performing Learners:} For learners who generate complex grammatical structures and long-form inquiries, the system must incorporate a technically robust Natural Language Understanding (NLU) engine capable of handling multi-clause utterances, disfluencies, and self-corrections without constraining deep exploration.
\end{itemize}

Third, an \textbf{Adaptive Timing} mechanism is required to regulate the timing of learner questions by jointly considering presentation modality and learner characteristics. Our findings suggest that heightened presence functions as a cognitive spotlight, optimizing task engagement by suppressing questioning during the action stage while facilitating inquiry during the ending stage. Moreover, individual learner characteristics—such as goal orientation and effort regulation—significantly shaped temporal questioning patterns. This suggests that question timing is shaped by both environmental and dispositional factors. To support these patterns, the following module is required:

\begin{itemize}
\item \textbf{Non-intrusive Bookmarking \& Post-Session Review:} The system should allow learners to log questions instantly via non-intrusive bookmarking without disrupting task flow. After the session, a post-session review dashboard should enable learners to revisit these entries and receive elaborated feedback, supporting both focused execution and reflective verification.
\end{itemize}

\subsection{Limitations and Future Work}
This study has several limitations. First, it was conducted in a controlled laboratory setting with a specific learning task (CPR), and the context of use and sample diversity were limited. These factors may restrict the generalizability of our findings to other domains or real-world educational contexts. 
Second, the analysis focused on interactions over a relatively short duration, which may have captured primarily short-term behavioral patterns influenced by initial learning stages or novelty effects.

Third, this study did not include the spatial dimension of questioning behavior. The ``Where'' aspect refers to the location of a learner’s visual attention at the moment a question is asked, typically measured through eye-tracking data. Because attentional targets were largely identical across conditions, this dimension was excluded from the present analysis. Fourth, question-asking behavior served as a narrow proxy for learner engagement. As questioning reflects only one facet of the broader, multi-dimensional construct of engagement, relying solely on verbal inquiry provides a limited view of learners’ overall engagement profiles.

To address these limitations, future work should examine these dynamics in more diverse and authentic learning scenarios. Longitudinal studies in real classrooms or in-the-field training contexts would be valuable for understanding how questioning strategies evolve as learners gain familiarity with AR technologies. In addition, incorporating eye-tracking would enable deeper analysis of the relationship between visual attention and verbal inquiry when attentional targets vary. Future research should also integrate broader engagement measures beyond question-asking to more comprehensively capture learners’ interaction patterns.

Moreover, another important avenue for future work is to design Adaptive IVAs that leverage our findings to provide tailored support by recognizing a learner’s motivational profile and information-seeking strategy in real time. For instance, a system could prompt learners to engage in conceptual deepening when their inquiry remains overly focused on procedural refinement, or adjust its feedback and encouragement strategies to align with learners’ motivational traits. Such adaptive support would contribute to optimizing individual learning experiences and maximizing the educational effectiveness of immersive technologies.

\section{Conclusion}
This study proposed a 5W1H-based framework to examine how presentation modality shapes learner questioning with IVAs in CPR instruction. Our findings reveal that the AR-IVA promoted more frequent and longer questions oriented toward conceptual deepening, whereas the Video-IVA encouraged procedural refinement. AR-IVA also heightened presence, which acted as a selective cognitive filter. Specifically, it suppressed questioning during action stages and increased inquiry during the ending stage. It also directed learners’ attention away from peripheral topics and toward the core concept. Furthermore, motivational and strategic traits moderated these effects by shaping the timing, content, intention, and intensity of questions. Overall, this study demonstrates that presentation modality influences questioning behavior through its interaction with learner characteristics, underscoring the need for adaptive, learner-centered IVA design.

\begin{acks}
This work was supported by Institute of Information \& communications Technology Planning \& Evaluation (IITP) grant funded by the Korea government (MSIT) (No. RS-2024-00397663, Real-time XR Interface Technology Development for Environmental Adaptation).
This work was supported by the IITP(Institute of Information \& Communications Technology Planning \& Evaluation)-ITRC(Information Technology Research Center) grant funded by the Korea government(Ministry of Science and ICT)(IITP-2026-RS-2024-00436398).
This research was supported by the MSIT(Ministry of Science and ICT), Korea, under the Graduate School of Metaverse Convergence support program(IITP-2022(2026)-RS-2022-00156435) supervised by the IITP(Institute for Information \& Communications Technology Planning \& Evaluation). 

\end{acks}

\bibliographystyle{ACM-Reference-Format}
\bibliography{template}










\end{document}